\documentclass[twocolumn,10pt,conference]{IEEEtran}
\usepackage[nolist,nohyperlinks]{acronym} 
\usepackage{algorithm,algpseudocode}
\usepackage{amsmath,amssymb,amsfonts}
\usepackage{array,booktabs,multirow,longtable}
\usepackage{graphicx,caption,subcaption,xspace,xcolor,cite}

\pdfoutput=1

\IEEEoverridecommandlockouts
\def\BibTeX{{\rm B\kern-.05em{\sc i\kern-.025em b}\kern-.08em  
    T\kern-.1667em\lower.7ex\hbox{E}\kern-.125emX}}

\usepackage{tikz,tikz-3dplot,tkz-tab}
\usetikzlibrary{angles, calc, positioning, quotes, intersections,fit,decorations.pathreplacing,decorations.markings}
\tdplotsetmaincoords{70}{120}

\tikzset{
    cross/.style={fill=white,path picture={\draw[black]
        (path picture bounding box.south east) -- (path picture bounding box.north west)
        (path picture bounding box.south west) -- (path picture bounding box.north east);}},
    dressed/.style={fill=white,postaction={pattern=north east lines}},
    momentum/.style={->,semithick,yshift=5pt,shorten >=5pt,shorten <=5pt},
    loop/.style 2 args={thick,decoration={markings,mark=at position {#1} with {\arrow{<},\node[anchor=\pgfdecoratedangle-90,font=\footnotesize] {};}},postaction={decorate}},
    label/.style={thin,gray,shorten <=-1.5ex}
}

\tikzset{
    cross/.pic = {
    \draw[rotate = 45] (-#1,0) -- (#1,0);
    \draw[rotate = 45] (0,-#1) -- (0, #1);
    }
}

\begin{document}
    \bstctlcite{IEEE:BSTcontrol}
    \begin{acronym}
        \acro{ALS}{alternating least squares}
        \acro{AoA}{azimuth of arrival}
        \acro{AoD}{azimuth of departure}
        \acro{AWGN}{additive white Gaussian noise}
        \acro{BALS}{bilinear alternating least squares}
        \acro{BD-RIS}{beyond-diagonal RIS}
        \acro{CRLB}{Cramér-Rao lower bound}
        \acro{CSI}{channel state information}
        \acro{DFRC}{dual-function radar-communication}
        \acro{DI}{Doppler ignorant}
        \acro{EoA}{elevation of arrival}
        \acro{EoD}{elevation of departure}
        \acro{ESPRIT}{estimation of the signal parameters via rotational invariance techniques}
        \acro{EVD}{eigenvalue decomposition}
        \acro{FIM}{Fisher information matrix}
        \acro{HOSVD}{higher-order singular value decomposition}
        \acro{ISAC}{integrated sensing and communications}
        \acro{KF}{Kronecker factorization}
        \acro{KRSA}{Khatri-Rao sum approximation}
        \acro{KSA}{Kronecker sum approximation}
        \acro{LOS}{line-of-sight}
        \acro{LS}{least squares}
        \acro{MIMO}{multiple-input multiple-output}
        \acro{ML}{maximum likelihood}
        \acro{NLOS}{non-line-of-sight}
        \acro{NMSE}{normalized mean squared error}
        \acro{NTFE}{nested tensor factorization estimation}
        \acro{OFDM}{orthogonal frequency division multiplexing}
        \acro{OMP}{orthogonal matching pursuit}
        \acro{PDF}{probability density function}
        \acro{RCS}{radar cross section}
        \acro{RIS}{reconfigurable intelligent surface}
        \acro{RMSE}{root mean squared error}
        \acro{SNR}{signal-to-noise ratio}
        \acro{SR}{sensing receiver}
        \acro{ST}{sensing transmitter}
        \acro{SVD}{singular value decomposition}
        \acro{TenDAE}{tensor Doppler-delay and angle estimation}
    \end{acronym}

    \title{Multi-Target Estimation via Tensor Decomposition for Beyond Diagonal RIS-Aided Bistatic Sensing
    \\
    }
    
    \author{\IEEEauthorblockN{Kenneth Benício,  
    André L. F. de Almeida, Fazal-E Asim, Bruno Sokal, \\
    Gabor Fodor, Behrooz Makki, and A. Lee Swindlehurst}}
    
    \maketitle

    \begin{abstract}
        We investigate the performance of beyond-diagonal reconfigurable intelligent surfaces (BD-RIS) for bistatic MIMO multi-target sensing using a two-stage tensor Doppler-delay-angle estimation (TenDAE). The first stage solves a Kronecker sum approximation (KSA) with a rank equal to the number of targets. The second stage employs a nested tensor factorization estimation (NTFE) that exploits the inherent multidimensional structure via two tensor decompositions that are solved in parallel. The first employs a PARAFAC decomposition to extract the targets' angles, and the second uses a nested PARAFAC decomposition to find the targets' delay and Doppler parameters. This two-stage approach decouples acquisition of the angles and delays/Dopplers using either alternating least squares or a higher-order singular value decomposition, followed by a high-resolution subspace technique, such as ESPRIT. We further compare the performance of a BD-RIS with a classical diagonal RIS. For the latter, we solve a Khatri-Rao sum approximation problem rather than the KSA due to the specific structure of the received signal. Notably, our NTFE framework remains blind to the underlying RIS architecture while simultaneously estimating all targets with minimal sensing resources. Additionally, we show that employing a nested-PARAFAC decomposition enables the decoupling of the delay-Doppler and angle domains. We also derive the Cramér-Rao lower bound to further assess the performance of the TenDAE framework. Finally, we numerically evaluate the solutions presented in this paper and demonstrate their efficiency in terms of RMSE compared with state-of-the-art approaches.
\end{abstract}
    

    \section{Introduction} \label{sec:introduction}
        \IEEEPARstart{O}{ver} the past decade, \ac{RIS} technology has emerged as a transformative approach for dynamically shaping wireless propagation environments for enhanced coverage, spectral efficiency, and \ac{SNR} \cite{Rui_Zhang_2021, zheng2022survey}. A typical passive \ac{RIS} consists of a planar array of reflecting elements whose phase shifts can be individually tuned \cite{gong2020toward}. However, optimal tuning requires estimation of the cascaded channel using pilot signals reflected by the \ac{RIS} under a prescribed training protocol. Several works have addressed this problem \cite{Swindle2022,de2021channel,deAraujoSAM2020,benicio2023tensor_wcl}. More recently, interest has shifted to leveraging passive \acs{RIS} to enhance sensing and localization at the \ac{ST}, motivating \ac{ISAC} architectures that improve both communication and sensing performance \cite{Fan_2022, 10243495}. Recent work has also explored novel ``beyond diagonal'' RIS (BD-RIS) architectures that introduce controllable inter-element coupling, yielding a generally non-diagonal phase-shift matrix \cite{9437234, 10352433, 9491943}. The resulting additional degrees of freedom can improve performance compared to classical diagonal \acs{RIS} designs \cite{li2023reconfigurable,li2022beyond}. \ac{BD-RIS} modeling and hardware architectures were first investigated in \cite{nerini2024beyond}. Following \cite{nerini2024beyond}, we adopt a fully-connected \ac{BD-RIS} model and focus on its use for multi-target parameter estimation; related applications to estimation and beamforming have also been considered in \cite{wang2024dual, liu2024enhancing, ginige2025efficient, de2025channel}.
        
        \indent \ac{BD-RIS} architectures for sensing/communication co-design have recently been proposed. In \cite{wang2024dual}, a \ac{BD-RIS}-aided dual-function radar-communication system is optimized to improve both communication and sensing performance. A fully-connected \ac{BD-RIS} is considered in \cite{liu2024enhancing} for \ac{ISAC}, demonstrating throughput gains while maintaining sensing gains relative to a diagonal \ac{RIS}. The work in \cite{ginige2025efficient} addresses parameter estimation and channel prediction by iteratively estimating decoupled \ac{BD-RIS} channels using a Tucker decomposition, yielding low-complexity estimates that initialize a prediction stage. Finally, \cite{de2025channel} proposes a block Tucker decomposition to improve estimation relative to conventional \ac{LS} methods and reduce training overhead. 
    
        \indent Building on these advancements, our previous work has addressed parameter estimation in diagonal-\ac{RIS}-assisted (single-connected) \ac{MIMO} sensing by exploiting the multidimensional structure of the received signal at the \ac{ST} \cite{tschudin1999comparison,feng2017comparison, benicio2024low, benicio2024ris}. In particular, we proposed a two-stage nested tensor approach, followed by \ac{ESPRIT}-based extraction of delay, Doppler, and angle parameters for a single target in \cite{benicio2024low, benicio2024ris}. 
        
        In this paper, we study multi-target estimation in BD-RIS-assisted networks. Our approach first estimates the rank-$K$ \ac{KSA} of the received signal, and then solves two third-order tensor models in parallel to jointly recover each target's delay, Doppler, and angle parameters. The proposed \ac{TenDAE} framework consists of two stages. In the first stage, a \ac{KSA} is employed to obtain low-rank subspace estimates of the effective channel. In the second, an \ac{NTFE} procedure jointly estimates the tensor factors that encapsulate the target parameters. The parameters of interest are recovered through a an extraction step using state-of-the-art estimation techniques. We also discuss identifiability and uniqueness conditions, and compare the fully connected \ac{BD-RIS} with its single-connected (diagonal) counterpart. We also compare the performance of our proposed algorithms with state-of-the-art methods, e.g., \cite{ercan2025ris}. Our results highlight the potential of fully-connected \ac{BD-RIS} architectures in enhancing multi-target sensing while enabling efficient tensor-based processing.

        The contributions of this paper are summarized as follows:

        \textit{First,} we show that target estimation using the signal reflected by a fully-connected \ac{BD-RIS} can be recast as a rank-$K$ \ac{KSA} that can be solved via a single \ac{SVD} to estimate the overall channel subspace. We then exploit (i) a third-order PARAFAC model to estimate the angle parameters and (ii) a third-order nested PARAFAC model to estimate the delay--Doppler parameters in parallel.

        \textit{Second,} we introduce the \ac{TenDAE} framework, which employs a modular two-stage process that combines subspace estimation with tensor fitting to achieve high accuracy and reduce computational complexity. In TenDAE, tensor models are solved via \ac{ALS}, with the final parameters extracted using \ac{ESPRIT}. We further extend the \ac{HOSVD}-based single-target solution from our previous work \cite{benicio2024low} to a general multi-target solution integrated into the \ac{TenDAE} framework. Our approach estimates all targets simultaneously, demonstrates robustness to noise, and requires minimal sensing resources.

        \textit{Third,} we propose a nested-PARAFAC tensor model that decouples the Doppler and angle domains with very low computational cost. This decoupling enhances estimation performance in both domains by enabling accurate acquisition of high-speed target data without compromising angle estimation due to Doppler-related imperfections.
        
        \textit{Fourth,} we compare fully-connected \ac{BD-RIS} and diagonal (single-connected) \ac{RIS} architectures. In particular, we show how the \ac{KSA} model induced by \ac{BD-RIS} reduces to a \ac{KRSA} model under a diagonal \ac{RIS}, and we discuss the corresponding identifiability implications.

        \textit{Fifth,} we study identifiability and uniqueness conditions for the proposed tensor models and derive system design guidelines that guarantee parameter recovery. In addition, we provide sufficient uniqueness conditions based on Kruskal's criteria for $N$th-order PARAFAC models.

        \textit{Sixth,} we derive the \ac{FIM} and associated \ac{CRLB} for the proposed scenario. The resulting bound characterizes the achievable estimation performance for the delay, Doppler, complex channel coefficients, and angle parameters of all $K$ targets.

        \textit{Finally,} we present numerical results demonstrating the advantage of the proposed tensor-based estimators over recent state-of-the-art methods \cite{kemal2024ris,ercan2025ris}, highlighting performance trade-offs and their proximity to the derived \ac{CRLB}.
        
        This work is organized as follows. Section II presents notation and tensor decomposition preliminaries. Section III presents the system model and formulates the \ac{KSA} problem for fully-connected \ac{BD-RIS} and the \ac{KRSA} problem for diagonal \ac{RIS}. Section IV details the received signal model. Section V develops the proposed parameter estimation method for multi-target sensing based on subspace estimation. Section VI discusses identifiability and uniqueness and their implications for system design. Section VII derives the \ac{CRLB}. Section IX presents numerical results, and Section X concludes the paper.
        
    \section{Tensor Preliminaries} \label{sec:tensor_preliminaries}
        \subsection{Notation and Properties}
            Scalars are represented in lower-case ($a$), column vectors in boldface lower-case ($\boldsymbol{a}$), matrices in boldface upper-case ($\boldsymbol{A}$), and tensors in calligraphic upper-case ($\boldsymbol{\mathcal{A}}$). The symbols $\{\cdot\}^{*}$, $\{\cdot\}^{\text{T}}$, $\{\cdot\}^{\text{H}}$, and $\{\cdot\}^{\dagger}$ represent the conjugate, transpose, conjugate transpose, and pseudoinverse operations, respectively. The matrix $\boldsymbol{I}_{N}$ denotes an identity matrix of dimension $N$. The operator $||\cdot||_{\text{F}}$ denotes the Frobenius norm of a matrix or tensor, and $\mathbb{E}\{\cdot\}$ is the expectation operator. The $j$th column of matrix $\boldsymbol{A} \in \mathbb{C}^{I \times R}$ is denoted by $\boldsymbol{a}_{j} \in \mathbb{C}^{I \times 1}$, and the operator $\text{D}_j(\boldsymbol{A})$ forms an $R\times R$ diagonal matrix from $\boldsymbol{a}_{j}$. The operator $\text{vec}(\boldsymbol{A})$ converts a matrix $\boldsymbol{A} \in \mathbb{C}^{I \times R}$ into a column vector $\boldsymbol{a} \in \mathbb{C}^{I R \times 1}$ by stacking its columns, while $\text{unvec}_{I \times R}(\boldsymbol{a})$ returns the vector to the original matrix $\boldsymbol{A} \in \mathbb{C}^{I \times R}$. Element-wise division is indicated by $\oslash$, and $\circ$, $\otimes$, and $\diamond$ represent the outer, Kronecker, and Khatri-Rao products, respectively. The operator $\sqcup_{n}$ indicates a concatenation along the $n$th dimension. We also employ the following properties throughout the paper:
            \begin{align}
                \text{vec}(\boldsymbol{a} \boldsymbol{b^{\text{T}}}) &= \boldsymbol{b} \otimes \boldsymbol{a}, \label{eq:property_1} \\
                \text{vec}(\boldsymbol{A} \boldsymbol{B} \boldsymbol{C}) &= (\boldsymbol{C}^{\text{T}} \otimes \boldsymbol{A}) \text{vec}(\boldsymbol{B}), \label{eq:property_2} \\
                \text{vec}(\boldsymbol{A} \text{D}(\boldsymbol{b}) \boldsymbol{C}) &= (\boldsymbol{C}^{\text{T}} \diamond \boldsymbol{A}) \boldsymbol{b}, \label{eq:property_3} \\
                (\boldsymbol{A} \otimes \boldsymbol{B}) (\boldsymbol{C} \otimes \boldsymbol{D}) &= (\boldsymbol{A} \boldsymbol{C}) \otimes (\boldsymbol{B} \boldsymbol{D}),  \label{eq:property_4} \\
                (\boldsymbol{A} \otimes \boldsymbol{B}) (\boldsymbol{C} \diamond \boldsymbol{D}) &= (\boldsymbol{A} \boldsymbol{C}) \diamond (\boldsymbol{B} \boldsymbol{D}).  \label{eq:property_5} 
            \end{align}
            For $\boldsymbol{C} = \boldsymbol{A} \otimes \boldsymbol{B} \in \mathbb{C}^{I_{2} I_{1} \times J_{2} J_{1}}$, where $\boldsymbol{A} \in \mathbb{C}^{I_{1} \times J_{1}}$ and $\boldsymbol{B} \in \mathbb{C}^{I_{2} \times J_{2}}$, the rearrangement operator that maps the original matrix $\boldsymbol{C}$ as $I_{1} J_{1}$ blocks of the same size $I_{2} \times J_{2}$ is
            \begin{align}
                \mathcal{R}_{I_{1} \times J_{1}}(\boldsymbol{C}) = \left[\text{vec}(\boldsymbol{C}^{I_{2}J_{2}}_{1,1}), \cdots, \text{vec}(\boldsymbol{C}^{I_{2}J_{2}}_{I_{1},J_{1}})\right]^{\text{T}},
            \end{align}
            where $\boldsymbol{C}^{I_{2}J_{2}}_{i_{1},j_{1}}$ is the $(i_{1}, j_{1})$th block of size $I_{2} \times J_{2}$.
        \subsection{Slices and Unfoldings}
            Consider a set of matrices $\boldsymbol{Y}_{r} \in \mathbb{C}^{I \times J}, \forall r \in \{1, \cdots, R\}$. By concatenating all the $R$ matrices, we build a $3$rd order tensor defined as $\boldsymbol{\mathcal{Y}} = \boldsymbol{Y}_{1} \sqcup_{3} \cdots \sqcup_{3} \boldsymbol{Y}_{R} \in \mathbb{C}^{I \times J \times R}$. Thus, the original matrix $\boldsymbol{Y}_{r}$ can be viewed as the $r$th frontal slice of the tensor $\boldsymbol{\mathcal{Y}}$ with notation $\boldsymbol{\mathcal{Y}}_{..r} = \boldsymbol{Y}_{r} \in \mathbb{C}^{I \times J}$. The frontal slice is obtained by varying the dimensions of the rows and columns for a fixed $3$rd dimension index $r$. Note that from the original $\boldsymbol{\mathcal{Y}}$ tensor, one can construct three matrices, each representing a different tensor slice. These matrices are referred to as the $n$-mode unfoldings, which, in the case of a $3$rd order tensor, can be represented as a function of the frontal slices:
            \begin{align}
                \left[\boldsymbol{\mathcal{Y}}\right]_{(1)} &= \left[\boldsymbol{\mathcal{Y}}_{..1}, \cdots, \boldsymbol{\mathcal{Y}}_{..R}\right] \in \mathbb{C}^{I \times J R}, \\
                \left[\boldsymbol{\mathcal{Y}}\right]_{(2)} &= \left[\boldsymbol{\mathcal{Y}}^{\text{T}}_{..1}, \cdots, \boldsymbol{\mathcal{Y}}^{\text{T}}_{..R}\right] \in \mathbb{C}^{J \times I R}, \\
                \left[\boldsymbol{\mathcal{Y}}\right]_{(3)} &= \left[\text{vec}(\boldsymbol{\mathcal{Y}}_{..1}), \cdots, \text{vec}(\boldsymbol{\mathcal{Y}}_{..R})\right] \in \mathbb{C}^{R \times I J}.
            \end{align}
            From the concept of unfolding, we can define the $n$-mode product operator, denoted by $\times_{n}$, that represents the product between a tensor $\boldsymbol{\mathcal{Y}}$ and a matrix $\boldsymbol{A}$, resulting in a tensor $\boldsymbol{\mathcal{X}}$ with compatible dimensions, i.e., $\boldsymbol{\mathcal{X}} = \boldsymbol{\mathcal{Y}} \times_{n} \boldsymbol{A}$. This operation can be accomplished by multiplying the $n$-mode unfolding of tensor $\boldsymbol{\mathcal{Y}}$ by the matrix $\boldsymbol{A}$, i.e., $\left[\boldsymbol{\mathcal{X}}\right]_{(n)} = \boldsymbol{A} \left[\boldsymbol{\mathcal{Y}}\right]_{(n)}$.
        \subsection{Tensor Decomposition}
            The Tucker decomposition represents a $3$rd order tensor $\boldsymbol{\mathcal{Y}} \in \mathbb{C}^{I \times J \times K}$ as the sum of multiple rank-one tensors:
            \begin{align}
                \boldsymbol{\mathcal{Y}} &= \sum^{R_{1}}_{r_{1} = 1} \sum^{R_{2}}_{r_{2} = 1} \sum^{R_{3}}_{r_{3} = 1} g_{r_{1} r_{2} r_{3}} \boldsymbol{a}_{r_{1}} \circ \boldsymbol{b}_{r_{2}} \circ \boldsymbol{c}_{r_{3}}
            \end{align}
            where $\boldsymbol{a}_{r_{1}} \in \mathbb{C}^{I \times 1}$, $\boldsymbol{b}_{r_{2}} \in \mathbb{C}^{J \times 1}$, and $\boldsymbol{c}_{r_{3}} \in \mathbb{C}^{K \times 1}$ are the column vectors of the factor matrices $\boldsymbol{A} \in \mathbb{C}^{I \times R_{1}}$, $\boldsymbol{B} \in \mathbb{C}^{J \times R_{2}}$, and $\boldsymbol{C} \in \mathbb{C}^{K \times R_{2}}$, respectively. We can also define the core tensor of the Tucker decomposition as $\boldsymbol{\mathcal{G}} \in \mathbb{C}^{R_{1} \times R_{2} \times R_{3}}$ with the typical element defined as $\left[\boldsymbol{\mathcal{G}}\right]_{r_{1} r_{2} r_{3}} = g_{r_{1} r_{2} r_{3}}$. The Tucker decomposition can be written as
            \begin{align}
                \boldsymbol{\mathcal{Y}} = \boldsymbol{\mathcal{G}} \times_{1} \boldsymbol{A} \times_{2} \boldsymbol{B} \times_{3} \boldsymbol{C}. \label{eq:tucker_decomposition}
            \end{align}
            \indent Letting $R_{1} = R_{2} = R_{3} = R$ and reducing the core tensor to the identity, we express the PARAFAC decomposition as the following special case of the Tucker decomposition:
            \begin{align}
                \boldsymbol{\mathcal{Y}} = \boldsymbol{\mathcal{I}}_{3,R} \times_{1} \boldsymbol{A} \times_{2} \boldsymbol{B} \times_{3} \boldsymbol{C}, \label{eq:parafac_decomposition}
            \end{align}
            where $\boldsymbol{A} \in \mathbb{C}^{I \times R}$, $\boldsymbol{B} \in \mathbb{C}^{J \times R}$, and $\boldsymbol{C} \in \mathbb{C}^{K \times R}$ are the factor matrices and $R$ represents the tensor rank. The tensor decomposition defined in (\ref{eq:parafac_decomposition}) is used throughout this work as a starting point for sensing a cluster of $K$ targets. Additionally, from the PARAFAC definition in (\ref{eq:parafac_decomposition}), we can define the nested PARAFAC decomposition, where one of the modes of the regular PARAFAC tensor is an unfolding of another PARAFAC tensor, and is expressed as 
            \begin{align}
                \boldsymbol{\mathcal{Y}} = \boldsymbol{\mathcal{I}}_{3,R} \times_{1} \boldsymbol{A} \times_{2} \boldsymbol{B} \times_{3} \left[\boldsymbol{\mathcal{X}}\right]_{(1)}, \label{eq:nested_parafac_decomposition}
            \end{align}
            where $\boldsymbol{\mathcal{X}} = \boldsymbol{\mathcal{I}}_{3,R} \times_{1} \boldsymbol{C} \times_{2} \boldsymbol{D} \times_{3} \boldsymbol{E}$ is a $3$rd order nested tensor that also follows a PARAFAC decomposition.
    \section{System Model} \label{sec:system_model}
       We consider a downlink bistatic scenario where the \ac{ST} and the \ac{SR} are equipped with multiple antennas and assisted by a fully-connected \ac{BD-RIS} with $N = N_{z} N_{y}$ passive reflecting elements to sense $K$ targets. The direct \ac{LOS} between the \ac{ST} and the target cluster is assumed blocked, so that target reflections are observed only through a virtual \ac{BD-RIS}-assisted \ac{LOS} path. The \ac{ST} transmits \ac{OFDM} pilot signals and receives echo signals via the \ac{BD-RIS}. The \ac{ST} and \ac{SR} employ rectangular arrays with $L_{\text{ST}} = L_{1y} L_{1z}$ and $L_{\text{SR}} = L_{2y} L_{2z}$ antennas, respectively, and the \ac{ST} transmits an \ac{OFDM} pulse with $Q$ subcarriers and $M$ symbols. The pilot on subcarrier $q$ and symbol $m$ is denoted by $\boldsymbol{x}_{q,m} \in \mathbb{C}^{L_{\text{ST}} \times 1}$. The received pilot signal at the \ac{SR} is given by
        \begin{align}
            \begin{split}&\boldsymbol{y}_{q,m,t} = \sum^{K}_{k = 1} \alpha_{k} \underbrace{\boldsymbol{a}_{\text{sr}}(\phi^{k}_{\text{sr}},\theta^{k}_{\text{sr}}) \boldsymbol{b}^{\text{T}}_{\text{tx}}(\phi^{k}_{\text{ris}_{\text{D}}}, \theta^{k}_{\text{ris}_{\text{D}}})}_{\text{RIS-Cluster-SR}} \boldsymbol{S}_{t} \\ \hspace{-0.1cm} &\times \underbrace{\boldsymbol{b}_{\text{rx}}(\hspace{-0.05cm}\phi_{\text{ris}_{\text{A}}},\hspace{-0.05cm}\theta_{\text{ris}_{\text{A}}}\hspace{-0.05cm}) \boldsymbol{a}^{\text{T}}_{\text{st}}(\hspace{-0.05cm}\phi_{\text{st}},\hspace{-0.05cm}\theta_{\text{st}}\hspace{-0.05cm})}_{\text{ST-RIS}} \boldsymbol{x}_{q,m} [\boldsymbol{c}(\tau_{k})]_{q} [\boldsymbol{d}(\nu_{k})]_{m} \hspace{-0.1cm}+\hspace{-0.1cm} \boldsymbol{z}_{q,m,t},\hspace{-0.1cm}
            \end{split}
        \end{align}
        where $\boldsymbol{a}_{\text{st}}(\phi_{\text{st}},\theta_{\text{st}}) \in \mathbb{C}^{L_{\text{ST}} \times 1}$ is the array response at the \ac{ST}, $\phi_{\text{st}}$ is the azimuth angle of departure (AoD) and $\theta_{\text{st}}$ is the elevation angle of departure (EoD). Similarly, $\boldsymbol{b}(\phi_{\text{ris}_{\text{A}}}, \theta_{\text{ris}_{\text{A}}}) \in \mathbb{C}^{N \times 1}$ is the steering vector at the \ac{BD-RIS} towards the \ac{ST}, with $\phi_{\text{ris}_{\text{A}}}$ and $\theta_{\text{ris}_{\text{A}}}$ representing the azimuth angle of arrival (AoA) and elevation angle of arrival (EoA), respectively. On the target side, $\boldsymbol{b}_{\text{tx}}(\phi^{k}_{\text{ris}_{\text{D}}}, \theta^{k}_{\text{ris}_{\text{D}}}) \in \mathbb{C}^{N \times 1}$ is the steering vector from the \ac{BD-RIS} to the $k$th target, where $\phi^{k}_{\text{ris}_{\text{D}}}$ and $\theta^{k}_{\text{ris}_{\text{D}}}$ are the AoD and EoD, respectively. Finally, $\boldsymbol{a}_{\text{sr}}(\phi^{k}_{\text{sr}},\theta^{k}_{\text{sr}}) \in \mathbb{C}^{L_{\text{SR}} \times 1}$ is the array response at the \ac{SR}, where $\phi^{k}_{\text{sr}}$ is the AoA and $\theta^{k}_{\text{sr}}$ is the EoA of the $k$th target. Assuming half-wavelength element spacing, and defining the spatial frequencies
        $\mu_{\text{st}} =  \pi \text{sin}(\phi_{\text{st}}) \text{sin}(\theta_{\text{st}})$ and $\psi_{\text{st}} =  \pi \text{cos}(\phi_{\text{st}})$, the \ac{ST} array response in the y-z plane is
        \begin{align}
            \boldsymbol{a}(\mu_{\text{st}},\psi_{\text{st}}) = \boldsymbol{a}_{y}(\mu_{\text{st}}) \otimes \boldsymbol{a}_{z}(\psi_{\text{st}}) \in \mathbb{C}^{L_{\text{ST}} \times 1},
        \end{align}
        where
        \begin{align}
            \notag \boldsymbol{a}_{y}(\mu_{\text{st}}) &= \left[1,  e^{-j\mu_{\text{st}}}, \cdots, e^{-j  (L_{1y} - 1) \mu_{\text{st}}} \right]^{\text{T}} \in \mathbb{C}^{L_{1y} \times 1}, \\
            \notag \boldsymbol{a}_{z}(\psi_{\text{s}t}) &= \left[1,  e^{-j\psi_{\text{st}}}, \cdots, e^{-j  (L_{1z} - 1) \psi_{\text{st}}} \right]^{\text{T}} \in \mathbb{C}^{L_{1z} \times 1}.
        \end{align}
        \indent We can define the array response for the \ac{BD-RIS}, $\boldsymbol{b}_{\text{rx}}(\phi_{\text{ris}_{\text{A}}}, \theta_{\text{ris}_{\text{A}}})$ and $\boldsymbol{b}_{\text{tx}}(\phi^{k}_{\text{ris}_{\text{D}}}, \theta^{k}_{\text{ris}_{\text{D}}})$, and the \ac{SR}, $\boldsymbol{a}_{\text{sr}}(\phi^{k}_{\text{sr}},\theta^{k}_{\text{sr}})$, similarly. The \ac{BD-RIS} phase-shift matrix linked to the $t$th time-slot is represented by $\boldsymbol{S}_{t} \in \mathbb{C}^{N \times N}$, with $\boldsymbol{S}^{\text{H}}_{t} \boldsymbol{S}_{t} = \boldsymbol{I}_{N}$ \cite{shen2021modeling}. The frequency-domain steering vector of the $k$th target due to the time delay $\tau_{k}$ is $\boldsymbol{c}(\tau_{k}) = \left[1, \cdots, e^{-j 2 \pi (Q - 1) \Delta f \tau_{k}}\right]^{\text{T}} \in \mathbb{C}^{Q \times 1}$, whose $q$th element $\left[\boldsymbol{c}(\tau_{k})\right]_{q}$ is linked to the $q$-th \ac{OFDM} subcarrier. Furthermore, $\boldsymbol{d}(\nu_{k}) = \left[1, \cdots, e^{j 2 \pi (M - 1) T_{s} \nu_{k}}\right]^{\text{T}} \in \mathbb{C}^{M \times 1}$ is the time-domain steering vector corresponding to the Doppler shift $\nu_{k}$ of the $k$th target, and it is linked to the $m$th \ac{OFDM} symbol by the $m$th element $\left[\boldsymbol{d}(\nu_{k})\right]_{m} \in \mathbb{C}$. Finally, $\boldsymbol{z}_{q,m,t} \in \mathbb{C}^{L \times 1}$ is \ac{AWGN} on the $q$th subcarrier of the $m$th symbol and $t$th time-slot. 
        
        \begin{figure}[!t]
            \centering
            \includegraphics[width=0.975\columnwidth]{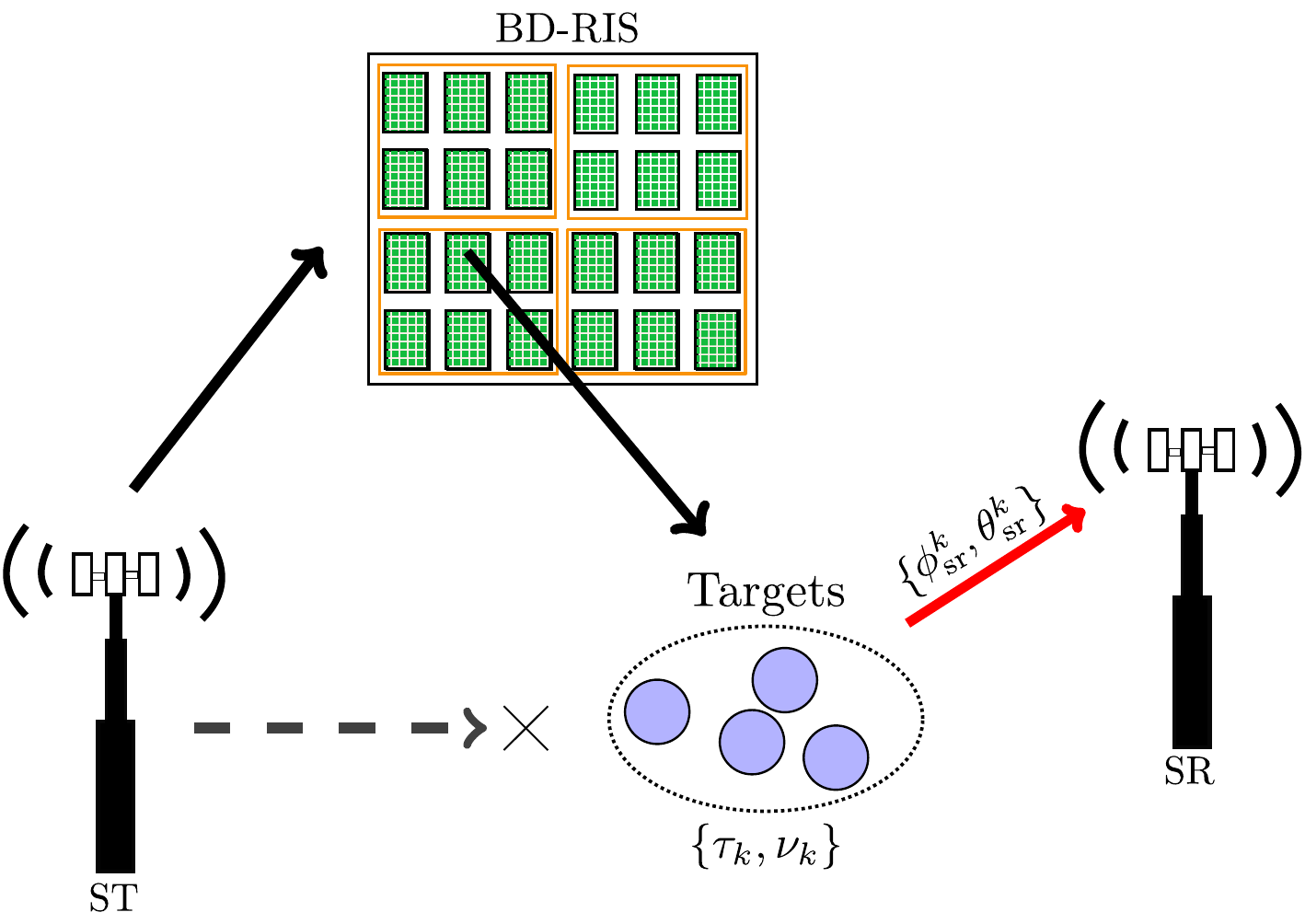}
            \caption{Bistatic BD-RIS-assisted multi-target localization.}
            \label{fig:system_model_localization}
        \end{figure}
        
\begin{figure}[!t]
            \centering
            \begin{tikzpicture}[scale=0.825, every node/.style={scale=0.65}]
                \begin{scope}[
                    box1/.style={draw=black, thick, rectangle, minimum height=0.5cm, minimum width=0.5cm}]
                    \draw[black,dashed,red,fill=white] (-4.65,-.85) rectangle (-2+0.35,.5);
                    \node[box1, fill=green!30] (c1) at (-4.25,0) {$\boldsymbol{S}_{1}$};
                    \node[box1, fill=green!30, right=.125cm of c1] (c2) {$\boldsymbol{S}_{1}$};
                    \node[right=.125cm of c2] (c3) {$\cdots$};
                    \node[box1, fill=green!30, right=.125cm of c3] (c4) {$\boldsymbol{S}_{1}$};
                    \draw[black,dashed,red,fill=white] (-1.50,-.85) rectangle (1.15+0.35,.5);
                    \node[box1, fill=green!30] (c1) at (-1.10,0) {$\boldsymbol{S}_{2}$};
                    \node[box1, fill=green!30, right=.125cm of c1] (c2) {$\boldsymbol{S}_{2}$};
                    \node[right=.125cm of c2] (c3) {$\cdots$};
                    \node[box1, fill=green!30, right=.125cm of c3] (c4) {$\boldsymbol{S}_{2}$};
                    \node[right=.275 of c4,circle,draw=black,fill=black,minimum size=1pt,scale=0.55] (T1) {};
                    \node[right=.1 of T1,circle,draw=black,fill=black,minimum size=1pt,scale=0.55] (T2) {};
                    \node[right=.1 of T2,circle,draw=black,fill=black,minimum size=1pt,scale=0.55] (T3) {};
                    \draw[black,dashed,red,fill=white] (2.35,-.85) rectangle (5.35,.5);
                    \node[box1, fill=green!30] (c1) at (2.75,0) {$\boldsymbol{S}_{T}$};
                    \node[box1, fill=green!30, right=.125cm of c1] (c2) {$\boldsymbol{S}_{T}$};
                    \node[right=.125cm of c2] (c3) {$\cdots$};
                    \node[box1, fill=green!30, right=.125cm of c3] (c4) {$\boldsymbol{S}_{T}$};
                    \draw[line width=0.15mm,black,<->] (-4.65,-1) -- (-1.65,-1) node[midway, below, rotate=+0]{\large Block size ($M Q$)};
                    \draw[line width=0.15mm,black,<->] (-4.65,0.75) -- (5.35,0.75) node[midway, above, rotate=+0]{\large Total of $T M Q$ OFDM symbols};
                    \draw[line width=0.15mm,black,<->] (-4.45,-0.5) -- (-2.25+0.35,-0.5) node[midway, below, rotate=+0]{$1$st block};
                    \draw[line width=0.15mm,black,<->] (-1.3,-0.5) -- (1+0.35,-0.5) node[midway, below, rotate=+0]{$2$nd block};
                    \draw[line width=0.15mm,black,<->] (2.5,-0.5) -- (4.85+0.35,-0.5) node[midway, below, rotate=+0]{$T$th block};
                \end{scope}
            \end{tikzpicture}
            \caption{Time-domain protocol. We assume the target parameters of interest do not vary during the estimation process. Also, the \ac{BD-RIS} remains constant over blocks of size $M Q$ and varies only between the $T$ blocks. 
            }
            \label{fig:transmission_protocol}
        \end{figure}
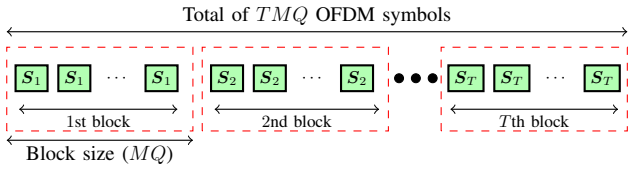

        Stacking the data from all $Q$ subcarriers and $M$ OFDM symbols, the received signal $\boldsymbol{Y}_{t} \in \mathbb{C}^{L \times M Q}$ is given by  
        \begin{align}
            \begin{split}
                &\boldsymbol{Y}_{t} = \sum^{K}_{k = 1} \boldsymbol{G}_{k} \boldsymbol{S}_{t} \boldsymbol{J}_{k} + \boldsymbol{Z}_{t} \in \mathbb{C}^{L_{\text{SR}} \times M Q}, \label{eq:received_signal_at_t}
            \end{split}
        \end{align}
        where $\boldsymbol{G}_{k} = \alpha_{k} \boldsymbol{a}_{\text{sr}}(\phi^{k}_{\text{sr}},\theta^{k}_{\text{sr}}) \boldsymbol{b}^{\text{T}}_{\text{tx}}(\phi^{k}_{\text{ris}_{\text{D}}}, \theta^{k}_{\text{ris}_{\text{D}}}) \in \mathbb{C}^{L_{\text{SR}} \times  N}$ is the channel between the \ac{BD-RIS} and the \ac{SR}, $\boldsymbol{J}_{k} = \boldsymbol{H} \boldsymbol{X} \text{D}(\boldsymbol{c}(\tau_{k}) \otimes \boldsymbol{d}(\nu_{k})) \in \mathbb{C}^{N \times M Q}$, $\boldsymbol{H} = \boldsymbol{b}_{\text{rx}}(\phi_{\text{ris}_{\text{A}}},\theta_{\text{ris}_{\text{A}}}) \boldsymbol{a}^{\text{T}}_{\text{st}}(\phi_{\text{st}},\hspace{-0.05cm}\theta_{\text{st}}) \in \mathbb{C}^{N \times L_{\text{ST}}}$ is the channel between the \ac{ST} and the \ac{BD-RIS}, and $\boldsymbol{X} \in \mathbb{C}^{L_{\text{ST}} \times M Q}$ is the transmitted pilot matrix. Applying the property $\text{vec}(\boldsymbol{A} \boldsymbol{B} \boldsymbol{C}) = (\boldsymbol
        {C}^{\text{T}} \otimes \boldsymbol{A}) \text{vec} (\boldsymbol{B})$ twice leads to 
        \begin{align}
            \notag \boldsymbol{y}_{t} &= \sum^{K}_{k = 1} \left(\boldsymbol{J}^{\text{T}}_{k} \otimes \boldsymbol{G}_{k}\right) \bar{\boldsymbol{s}}_{t} + \boldsymbol{z}_{t},
        \end{align}
        where $\boldsymbol{y}_{t} = \text{vec}(\boldsymbol{Y}_{t}) \in \mathbb{C}^{L_{\text{SR}} M Q \times 1}$, $\boldsymbol{z}_{t} = \text{vec}(\boldsymbol{Z}_{t}) \in \mathbb{C}^{L_{\text{SR}} M Q \times 1}$ and $\bar{\boldsymbol{s}}_{t} = \text{vec}(\boldsymbol{S}_{t}) \in \mathbb{C}^{N^{2} \times 1}$. Collecting all $T$ samples of the pilot signals as $\boldsymbol{Y} = \left[\boldsymbol{y}_{1}, \cdots, \boldsymbol{y}_{T}\right]$ and $\boldsymbol{Z} = \left[\boldsymbol{z}_{1}, \cdots, \boldsymbol{z}_{T}\right]$, we obtain
        \begin{align}
            \boldsymbol{Y} &= \sum^{K}_{k = 1} \left\{ \boldsymbol{J}^{\text{T}}_{k} \otimes \boldsymbol{G}_{k}\right\} \boldsymbol{S} + \boldsymbol{Z} \in \mathbb{C}^{L_{\text{SR}} M Q \times T}, \label{eq:received_signal_all_samples}
        \end{align}
        where $\boldsymbol{S} = \left[\bar{\boldsymbol{s}}_{1}, \cdots, \bar{\boldsymbol{s}}_{T}\right] \in \mathbb{C}^{N^{2} \times T}$. The proposed time-domain transmission protocol is shown in Fig.~\ref{fig:transmission_protocol}. 
        
        For the case of a diagonal \ac{RIS}, (\ref{eq:received_signal_all_samples}) becomes
        \begin{align}
            \boldsymbol{Y} &= \sum^{K}_{k = 1} \left\{ \boldsymbol{J}^{\text{T}}_{k} \diamond \boldsymbol{G}_{k}\right\} \boldsymbol{S}' + \boldsymbol{Z} \in \mathbb{C}^{L_{\text{SR}} M Q \times T}, \label{eq:received_signal_all_samples_dris}
        \end{align}
        where $\boldsymbol{S}' \in \mathbb{C}^{N \times T}$ is the phase-shift matrix of the diagonal \ac{RIS}. In the noiseless case, Eqs. (\ref{eq:received_signal_all_samples}) and (\ref{eq:received_signal_all_samples_dris}) are an exact \ac{KSA} and \ac{KRSA}, respectively, both of rank $K$.

    \section{Tensor Modeling Framework} \label{sec:tensor_modeling_framework}
       Here we derive a rank-$K$ Kronecker-sum approximation of the received signal after right-filtering the \ac{BD-RIS} phase-shift matrix in (\ref{eq:received_signal_all_samples}). We then exploit the resulting structure to construct two tensor models that can be processed in parallel: (i) a third-order PARAFAC tensor for estimating the angle parameters of the \text{RIS--cluster--SR} channel, and (ii) a third-order nested PARAFAC tensor that yields the \ac{ST}--\ac{RIS} arrival angles and the target delay--Doppler parameters.
        \begin{figure}[!t]
            \centering
            \includegraphics[width=0.975\columnwidth]{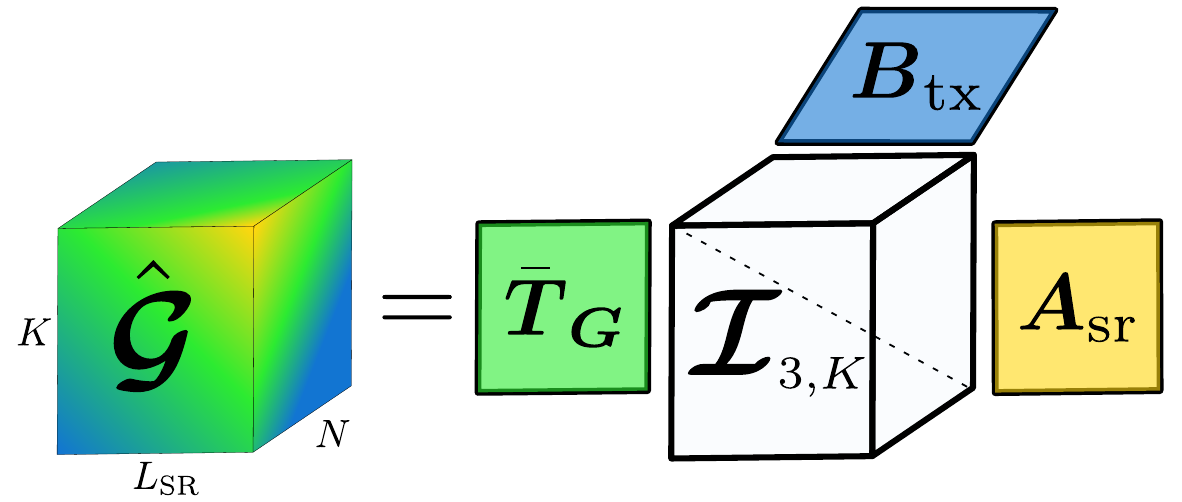}
            \caption{Illustration of the noiseless PARAFAC tensor $\hat{\boldsymbol{\mathcal{G}}}$ of dimensions $K \times L_{\text{SR}} \times N$. This tensor contains the information necessary to estimate the angles of the $K$ targets.}
            \label{fig:parafac}
        \end{figure}
            
            Starting from (\ref{eq:received_signal_all_samples}) and assuming $T \geq N^{2}$, we right-filter the phase-shift matrix to obtain
            \begin{align}
                \boldsymbol{Y}' &= \boldsymbol{Y} \boldsymbol{S}^{\dagger} \approx \sum^{K}_{k = 1} \left\{ \boldsymbol{J}^{\text{T}}_{k} \otimes \boldsymbol{G}_{k}\right\} \in \mathbb{C}^{L_{\text{SR}} M Q \times N^{2}}, \label{eq:received_signal_filtered}
            \end{align}
            and estimate the rank-$K$ \ac{KSA} factors by solving
            \begin{align}
                \{\hat{\boldsymbol{J}}_{k}, \hat{\boldsymbol{G}}_{k}\} = \underset{\boldsymbol{J}_{k}, \boldsymbol{G}_{k}}{\text{arg min}} \left|\left|\begin{aligned}\boldsymbol{Y}' -\sum^{K}_{k = 1} \left\{ \boldsymbol{J}^{\text{T}}_{k} \otimes \boldsymbol{G}_{k}\right\}\end{aligned}\right|\right|^{2}_{\text{F}}, \label{eq:ksa_1}
            \end{align}
            which provides the \ac{CSI} $\{\boldsymbol{J}^{\text{T}}_{k},\boldsymbol{G}_{k}\}_{k=1}^{K}$. Following \cite{cai2022kopa}, (\ref{eq:ksa_1}) can be recast as
            \begin{align}
                \{\hat{\boldsymbol{J}}, \hat{\boldsymbol{G}}\} &= \underset{\boldsymbol{J}, \boldsymbol{G}}{\text{arg min}} \left|\left|\begin{aligned}\mathcal{R}_{M Q \times N}(\boldsymbol{Y}') - \boldsymbol{G} \boldsymbol{
                J}^{\text{T}}\end{aligned}\right|\right|^{2}_{\text{F}}, \label{eq:ksa_2}
            \end{align}
            where $\boldsymbol{G} = \left[\text{vec}(\boldsymbol{G}_{1}), \cdots, \text{vec}(\boldsymbol{G}_{K})\right] \in \mathbb{C}^{L_{\text{SR}} N \times K}$, $\boldsymbol{J} = \left[\text{vec}(\boldsymbol{J}^{\text{T}}_{1}), \cdots, \text{vec}(\boldsymbol{J}^{\text{T}}_{K})\right] \in \mathbb{C}^{M Q N \times K}$, and $\mathcal{R}_{M Q \times N}(\boldsymbol{Y}')$ are rearrangements of the original $\boldsymbol{Y}'$, defined as
            \begin{align}
                \mathcal{R}_{M Q \times N}\left(\sum^{K}_{k = 1} \boldsymbol{J}^{\text{T}}_{k} \otimes \boldsymbol{G}_{k}\right) = \sum^{K}_{k = 1} \boldsymbol{g}_{k} \boldsymbol{j}^{\text{T}}_{k}. \label{eq:ksa_to_svd}
            \end{align}
            Eq.~(\ref{eq:ksa_to_svd}), which shows the connection between the \ac{KSA} and the \ac{SVD}, was first derived in \cite{pitsianis1993approximation}. The mapping $\mathcal{R}_{M Q \times N}: \mathbb{C}^{L_{\text{SR}} M Q \times N N} \rightarrow \mathbb{C}^{L_{\text{SR}} N \times M Q N}$ results in a matrix of rank $K$. Thus, to solve the problem in (\ref{eq:ksa_2}), we employ the \ac{SVD} $\mathcal{R}_{M Q \times N}(\boldsymbol{Y}')  = \boldsymbol{U} \boldsymbol{S} \boldsymbol{V}^{\text{H}}$ and define the estimates of the composite matrices containing the \ac{CSI} as $\hat{\boldsymbol{G}} = \boldsymbol{U}_{.1:K}$ and $\hat{\boldsymbol{J}} = \boldsymbol{V}^{*}_{.1:K}$, which are related to the true values as $\hat{\boldsymbol{G}} = \boldsymbol{G} \boldsymbol{T}_{\boldsymbol{G}}$ and $\hat{\boldsymbol{J}} = \boldsymbol{J} \boldsymbol{T}_{\boldsymbol{J}}$, for some transformations $\boldsymbol{T}_{\boldsymbol{G}}$ and $\boldsymbol{T}_{\boldsymbol{J}}$ due to the \textit{via} \ac{SVD} representation. 
            
            From these estimates, we can define two tensor models that will enable us to estimate the target parameters. From $\hat{\boldsymbol{G}}$ and applying Property (\ref{eq:property_1}), we obtain
            \begin{align}
                \notag \hat{\boldsymbol{G}} &= \left[\text{vec}(\boldsymbol{G}_{1}), \cdots, \text{vec}(\boldsymbol{G}_{K})\right] \boldsymbol{T}_{\boldsymbol{G}}, \\
                &= (\boldsymbol{B}_{\text{tx}} \diamond \boldsymbol{A}_{\text{sr}}) \boldsymbol{\Lambda}_{\boldsymbol{\alpha}}\boldsymbol{T}_{\boldsymbol{G}} \in \mathbb{C}^{L_{\text{SR}} N \times K}, \label{eq:G_1_unfold_transpose}
            \end{align}
            where $\boldsymbol{B}_{\text{tx}} = \left[\boldsymbol{b}^{\text{T}}_{\text{tx}}(\phi^{1}_{\text{ris}_{\text{D}}}, \theta^{1}_{\text{ris}_{\text{D}}}), \cdots, \boldsymbol{b}^{\text{T}}_{\text{tx}}(\phi^{K}_{\text{ris}_{\text{D}}}, \theta^{K}_{\text{ris}_{\text{D}}})\right] \in \mathbb{C}^{N \times K}$, $\boldsymbol{A}_{\text{sr}} = \left[\boldsymbol{a}_{\text{sr}}(\phi^{1}_{\text{sr}},\theta^{1}_{\text{sr}}), \cdots, \boldsymbol{a}_{\text{sr}}(\phi^{K}_{\text{sr}},\theta^{K}_{\text{sr}})\right] \in \mathbb{C}^{L_{\text{SR}} \times K}$, $\boldsymbol{\Lambda}_{\boldsymbol{\alpha}} = \text{D}(\boldsymbol{\alpha}) \in \mathbb{C}^{K \times K}$, and $\boldsymbol{\alpha} = \left[\alpha_{1}, \cdots, \alpha_{K}\right] \in \mathbb{C}^{K \times 1}$. Eq.~(\ref{eq:G_1_unfold_transpose}) is the first mode unfolding of a $3$rd order PARAFAC tensor, similar to (\ref{eq:parafac_decomposition}) as shown in Fig. \ref{fig:parafac}, given by
            \begin{align}
                \hat{\boldsymbol{\mathcal{G}}} &= \boldsymbol{\mathcal{I}}_{3,K} \times_{1} \bar{\boldsymbol{T}}_{\boldsymbol{G}} \times_{2} \boldsymbol{A}_{\text{sr}} \times_{3} \boldsymbol{B}_{\text{tx}} \in \mathbb{C}^{K \times L_{\text{SR}} \times N}, \label{eq:tensor_G_1}
            \end{align}
            whose unfoldings admit the following factorizations            
            \begin{align}
                [\hat{\boldsymbol{\mathcal{G}}}]_{(1)} &=  \bar{\boldsymbol{T}}_{\boldsymbol{G}} \left(\boldsymbol{B}_{\text{tx}} \diamond \boldsymbol{A}_{\text{sr}}\right)^{\text{T}}\in \mathbb{C}^{K \times L_{\text{SR}} N}, \label{eq:tensor_G_1_unfolding_1} \\
                [\hat{\boldsymbol{\mathcal{G}}}]_{(2)} &=  \boldsymbol{A}_{\text{sr}} \left(\boldsymbol{B}_{\text{tx}} \diamond \bar{\boldsymbol{T}}_{\boldsymbol{G}}\right)^{\text{T}} \in \mathbb{C}^{L_{\text{SR}} \times K N}, \label{eq:tensor_G_1_unfolding_2} \\
                [\hat{\boldsymbol{\mathcal{G}}}]_{(3)} &=  \boldsymbol{B}_{\text{tx}} \left(\boldsymbol{A}_{\text{sr}} \diamond \bar{\boldsymbol{T}}_{\boldsymbol{G}}\right)^{\text{T}}\in \mathbb{C}^{N \times K L_{\text{SR}}}, \label{eq:tensor_G_1_unfolding_3} 
            \end{align}
            where $\bar{\boldsymbol{T}}_{\boldsymbol{G}} = \boldsymbol{\Lambda}_{\boldsymbol{\alpha}}\boldsymbol{T}_{\boldsymbol{G}} \in \mathbb{C}^{K \times K}$. 
            
            Similarly, from $\hat{\boldsymbol{J}}$ we can define a nested PARAFAC tensor by applying Property (\ref{eq:property_2}) to obtain
            \begin{align}
                \notag \hat{\boldsymbol{J}} &= \left[\text{vec}(\boldsymbol{J}^{\text{T}}_{1}), \cdots, \text{vec}(\boldsymbol{J}^{\text{T}}_{K})\right] \boldsymbol{T}_{\boldsymbol{J}}  \\
                &= \left((\boldsymbol{H} \boldsymbol{X}) \diamond \boldsymbol{I}_{M Q}\right) \boldsymbol{F}_{\tau \nu} \boldsymbol{T}_{\boldsymbol{J}} \in \mathbb{C}^{M Q N \times K}, \label{eq:J_1_unfold_transpose}
            \end{align}
            where $\boldsymbol{F}_{\tau \nu} = \left[\boldsymbol{c}(\tau_{1}) \otimes \boldsymbol{d}(\nu_{1}) \cdots \boldsymbol{c}(\tau_{K}) \otimes \boldsymbol{d}(\nu_{K}) \right]^{\text{T}} = \left(\boldsymbol{C}_{\tau} \diamond \boldsymbol{D}_{\nu} \right)^{\text{T}} \in \mathbb{C}^{K \times M Q}$ is the matrix containing the coupled delay-Doppler information, $\boldsymbol{C}_{\tau} = \left[\boldsymbol{c}(\tau_{1}), \cdots, \boldsymbol{c}(\tau_{K})\right] \in \mathbb{C}^{Q \times K}$, $\boldsymbol{D}_{\nu} = \left[\boldsymbol{d}(\nu_{1}), \cdots, \boldsymbol{d}(\nu_{K})\right] \in \mathbb{C}^{M \times K}$, and $\boldsymbol{T}_{\boldsymbol{J}} \in \mathbb{C}^{K \times K}$. From (\ref{eq:J_1_unfold_transpose}), we define the transpose third mode unfolding of a nested PARAFAC tensor, similar to (\ref{eq:nested_parafac_decomposition}), as
            \begin{align}
                \hspace{-0.25cm} \hat{\boldsymbol{\mathcal{J}}} &= \boldsymbol{\mathcal{I}}_{3,M Q} \times_{1} \boldsymbol{I}_{M Q} \times_{2} \boldsymbol{H} \boldsymbol{X} \times_{3} \left[\boldsymbol{\mathcal{F}}\right]_{(1)} \hspace{-0.05cm}\in\hspace{-0.05cm} \mathbb{C}^{M Q \times N \times K}, \label{eq:tensor_J_1}
            \end{align}
            where $\left[\boldsymbol{\mathcal{F}}\right]_{(1)} = \boldsymbol{T}_{\boldsymbol{J}} \boldsymbol{F}^{\text{T}}_{\tau \nu} = \boldsymbol{T}_{\boldsymbol{J}} \left(\boldsymbol{C}_{\tau} \diamond \boldsymbol{D}_{\nu} \right)^{\text{T}}$ is the first mode unfolding of the PARAFAC tensor nested on the third mode of tensor $\boldsymbol{\mathcal{J}}$ whose unfoldings are given by
            \begin{align}
                [\hat{\boldsymbol{\mathcal{J}}}]_{(1)} &= \boldsymbol{I}_{M Q} \left(\left[\boldsymbol{\mathcal{F}}\right]_{(1)} \diamond \boldsymbol{H} \boldsymbol{X}\right)^{\text{T}}\in \mathbb{C}^{M Q \times N K}, \label{eq:tensor_J_1_unfolding_1} \\
                [\hat{\boldsymbol{\mathcal{J}}}]_{(2)} &= \boldsymbol{H} \boldsymbol{X} \left(\left[\boldsymbol{\mathcal{F}}\right]_{(1)} \diamond \boldsymbol{I}_{M Q}\right)^{\text{T}} \in \mathbb{C}^{N \times M Q K}, \label{eq:tensor_J_1_unfolding_2} \\
                [\hat{\boldsymbol{\mathcal{J}}}]_{(3)} &=  \left[\boldsymbol{\mathcal{F}}\right]_{(1)} \left(\boldsymbol{H} \boldsymbol{X} \diamond \boldsymbol{I}_{M Q}\right)^{\text{T}}\in \mathbb{C}^{K \times M Q N}. \label{eq:tensor_J_1_unfolding_3} 
            \end{align}
            The nested mode of $\hat{\boldsymbol{\mathcal{J}}}$ is a $3$rd order PARAFAC tensor:
            \begin{align}
                \boldsymbol{\mathcal{F}} &= \boldsymbol{\mathcal{I}}_{3,K} \times_{1} \boldsymbol{T}_{\boldsymbol{J}} \times_{2} \boldsymbol{D}_{\nu} \times_{3} \boldsymbol{C}_{\tau} \in \mathbb{C}^{K \times M \times Q}, \label{eq:tensor_F_1}
            \end{align}
            with unfoldings
            \begin{align}
                [\hat{\boldsymbol{\mathcal{J}}}]_{(1)} &= \boldsymbol{T}_{\boldsymbol{J}} \left(\boldsymbol{C}_{\tau} \diamond \boldsymbol{D}_{\nu}\right)^{\text{T}}\in \mathbb{C}^{K \times M Q}, \label{eq:tensor_F_1_unfolding_1} \\
                [\hat{\boldsymbol{\mathcal{J}}}]_{(2)} &= \boldsymbol{D}_{\nu} \left(\boldsymbol{C}_{\tau} \diamond \boldsymbol{T}_{\boldsymbol{J}}\right)^{\text{T}} \in \mathbb{C}^{M \times K Q}, \label{eq:tensor_F_1_unfolding_2} \\
                [\hat{\boldsymbol{\mathcal{J}}}]_{(3)} &=  \boldsymbol{C}_{\tau} \left(\boldsymbol{D}_{\nu} \diamond \boldsymbol{T}_{\boldsymbol{J}}\right)^{\text{T}}\in \mathbb{C}^{Q \times K M}. \label{eq:tensor_F_1_unfolding_3} 
            \end{align}
            \begin{figure}[!t]
                \centering
                \includegraphics[width=0.975\columnwidth]{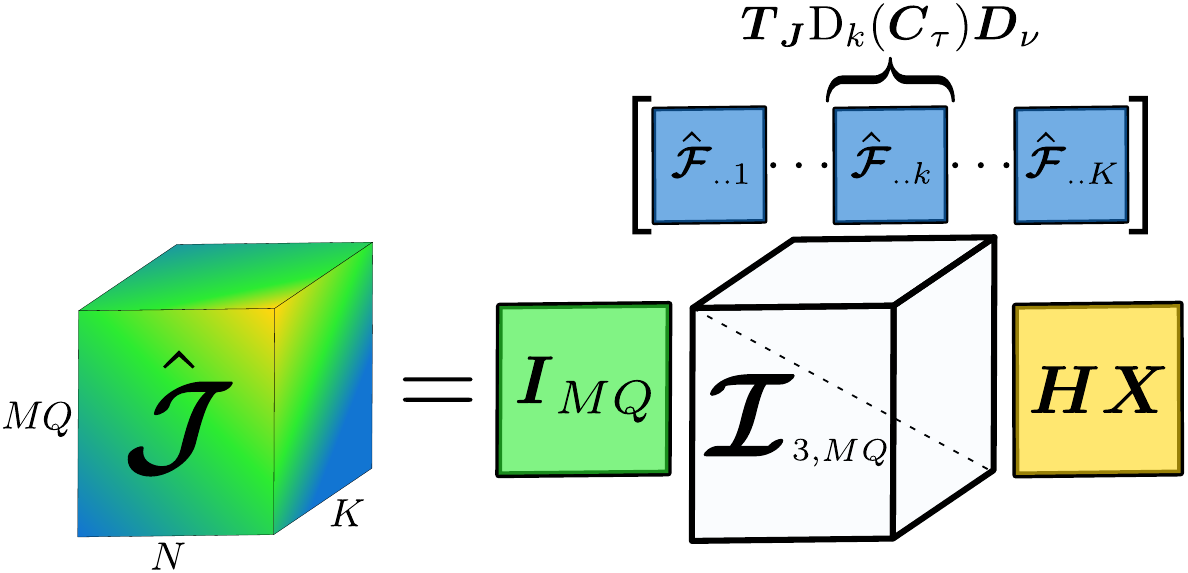}
                \caption{Illustration of the noiseless nested PARAFAC tensor $\hat{\boldsymbol{\mathcal{J}}}$ of dimension $M Q \times N \times K$. The $3$rd mode unfolding of $\hat{\boldsymbol{\mathcal{J}}}$, i.e., $[\hat{\boldsymbol{\mathcal{J}}}]_{(3)}$, is the nested mode encapsulating the $1$st mode unfolding of the tensor $\hat{\boldsymbol{\mathcal{F}}}$ of dimension $K \times M \times Q$.
                }
                \label{fig:nested_parafac}
            \end{figure}
   
    \section{Tensor-Based Multi-Target Sensing} \label{sec:tensor_based_multi_target_sensing}
        Here we present solutions for the tensor models in (\ref{eq:tensor_G_1}), (\ref{eq:tensor_J_1}), and (\ref{eq:tensor_F_1}). We fit these models with \ac{ALS} \cite{comon2009tensor, de2016overview}, which yields subspace estimates of the \ac{CSI} factors $\boldsymbol{G}$ and $\boldsymbol{J}$. The parameters of interest are then extracted via \ac{ESPRIT}, and the complex gains $\boldsymbol{\alpha}$ are obtained through a closed-form \ac{LS} step.

        \subsection{Cluster Angular Information Estimation}
            Starting from the PARAFAC model in (\ref{eq:tensor_G_1}), we estimate the target angles at the \ac{SR} through $\boldsymbol{A}_{\text{sr}}$ and the \ac{BD-RIS} departure angles through $\boldsymbol{B}_{\text{tx}}$ by solving
            \begin{equation}
                \hspace{-0.15cm}\left\{\hat{\boldsymbol{A}}_{\text{sr}}, \hat{\boldsymbol{B}}_{\text{tx}}\right\} \hspace{-0.075cm}=\hspace{-0.075cm} \underset{\boldsymbol{A}_{\text{sr}}, \boldsymbol{B}_{\text{tx}}}{\text{arg min}} \hspace{-0.075cm} \left| \hspace{-0.05cm} \left|\hat{\boldsymbol{\mathcal{G}}} \hspace{-0.075cm}-\hspace{-0.075cm} \boldsymbol{\mathcal{I}}_{3,K} \hspace{-0.075cm}\times_{1}\hspace{-0.075cm} \bar{\boldsymbol{T}}_{\boldsymbol{G}} \hspace{-0.075cm}\times_{2}\hspace{-0.075cm} \boldsymbol{A}_{\text{sr}} \hspace{-0.075cm}\times_{3}\hspace{-0.075cm} \boldsymbol{B}_{\text{tx}} \right| \hspace{-0.05cm}\right|^{2}_{\text{F}}.\label{eq:tensor_G_fit_problem_1}
            \end{equation}
            This problem is solved using \ac{ALS}, which alternates between solving the following optimization problems until a stopping criterion is reached:
            \begin{align}
                 \hat{\bar{\boldsymbol{T}}}_{\boldsymbol{G}} &= \underset{\bar{\boldsymbol{T}}_{\boldsymbol{G}}}{\text{arg min}} \left|\left|\begin{aligned}\left[\hat{\boldsymbol{\mathcal{G}}}\right]_{(1)} - \bar{\boldsymbol{T}}_{\boldsymbol{G}} \left(\boldsymbol{B}_{\text{tx}} \diamond \boldsymbol{A}_{\text{sr}}\right)^{\text{T}}&\end{aligned}\right|\right|^{2}_{\text{F}}, \label{eq:tensor_G_1_ls_1} \\
                &= \left[\hat{\boldsymbol{\mathcal{G}}}\right]_{(1)} \left[\left(\boldsymbol{B}_{\text{tx}} \diamond \boldsymbol{A}_{\text{sr}}\right)^{\text{T}}\right]^{\dagger}, \label{eq:tensor_G_1_pseudoinverse_1} \\
                 \hat{\boldsymbol{A}}_{\text{sr}} &= \underset{\boldsymbol{A}_{\text{sr}}}{\text{arg min}} \left|\left|\begin{aligned}\left[\hat{\boldsymbol{\mathcal{G}}}\right]_{(2)} - \boldsymbol{A}_{\text{sr}} \left(\boldsymbol{B}_{\text{tx}} \diamond \bar{\boldsymbol{T}}_{\boldsymbol{G}}\right)^{\text{T}}&\end{aligned}\right|\right|^{2}_{\text{F}}, \label{eq:tensor_G_1_ls_2} \\
                &= \left[\hat{\boldsymbol{\mathcal{G}}}\right]_{(2)} \left[\left(\boldsymbol{B}_{\text{tx}} \diamond \bar{\boldsymbol{T}}_{\boldsymbol{G}}\right)^{\text{T}}\right]^{\dagger}, \label{eq:tensor_G_1_pseudoinverse_2} \\
                 \hat{\boldsymbol{B}}_{\text{tx}} &= \underset{\boldsymbol{B}_{\text{tx}}}{\text{arg min}} \left|\left|\begin{aligned}\left[\hat{\boldsymbol{\mathcal{G}}}\right]_{(3)} - \boldsymbol{B}_{\text{tx}} \left(\boldsymbol{A}_{\text{sr}} \diamond \bar{\boldsymbol{T}}_{\boldsymbol{G}}\right)^{\text{T}}&\end{aligned}\right|\right|^{2}_{\text{F}}, \label{eq:tensor_G_1_ls_3} \\
                &= \left[\hat{\boldsymbol{\mathcal{G}}}\right]_{(3)} \left[\left(\boldsymbol{A}_{\text{sr}} \diamond \bar{\boldsymbol{T}}_{\boldsymbol{G}}\right)^{\text{T}}\right]^{\dagger}. \label{eq:tensor_G_1_pseudoinverse_3} 
            \end{align}
            The solutions above exist as long as $L_{\text{SR}} N \geq K$. 
            
            We then extract the $K$ target angles $(\phi^{k}_{\text{sr}},\theta^{k}_{\text{sr}})$ from $\hat{\boldsymbol{A}}_{\text{sr}} \in \mathbb{C}^{L_{\text{SR}} \times K}$ using $2$D \ac{ESPRIT}. Specifically, we compute the rank-$K$ \ac{SVD} $\hat{\boldsymbol{A}}_{\text{sr}} = \sum^{K}_{k = 1} \sigma_{k} \boldsymbol{u}_{k} \boldsymbol{v}^{\text{H}}_{k}$ and apply \ac{ESPRIT} to $\hat{\boldsymbol{a}}_{\text{sr}}(\phi^{k}_{\text{sr}},\theta^{k}_{\text{sr}}) = \sqrt{\sigma_{k}} \boldsymbol{u}_{k}$ to obtain $(\hat{\phi}^{k}_{\text{sr}},\hat{\theta}^{k}_{\text{sr}})$. Similarly, we compute $\hat{\boldsymbol{B}}_{\text{tx}} = \sum^{K}_{k = 1} \sigma'_{k} \boldsymbol{u}'_{k} \boldsymbol{v}^{'\text{H}}_{k}$ and apply $2$D \ac{ESPRIT} to $\hat{\boldsymbol{b}}_{\text{tx}}(\phi^{k}_{\text{ris}_{\text{D}}},\theta^{k}_{\text{ris}_{\text{D}}}) = \sqrt{\sigma'_{k}} \boldsymbol{u}'_{k}$ to estimate the \ac{BD-RIS} departure angles $(\hat{\phi}^{k}_{\text{ris}_{\text{D}}},\hat{\theta}^{k}_{\text{ris}_{\text{D}}})$. Finally, we refine the \ac{BD-RIS}--\ac{SR} factors by enforcing the array manifold structure
            \begin{align}
                \hat{\boldsymbol{A}}_{\text{sr}} &= \left[\boldsymbol{a}_{\text{sr}}(\hat{\phi}^{1}_{\text{sr}},\hat{\theta}^{1}_{\text{sr}}), \cdots, \boldsymbol{a}_{\text{sr}}(\hat{\phi}^{K}_{\text{sr}},\hat{\theta}^{K}_{\text{sr}})\right] \in \mathbb{C}^{L_{\text{SR}} \times K}, \\
                \hat{\boldsymbol{B}}_{\text{tx}} &= \left[\boldsymbol{b}_{\text{tx}}(\hat{\phi}^{1}_{\text{ris}_{\text{D}}},\hat{\theta}^{1}_{\text{ris}_{\text{D}}}), \cdots, \boldsymbol{b}_{\text{tx}}(\hat{\phi}^{K}_{\text{ris}_{\text{D}}},\hat{\theta}^{K}_{\text{ris}_{\text{D}}})\right] \in \mathbb{C}^{N \times K}.
            \end{align}
            The first stage is summarized in Algorithm~\ref{alg:proposed_1}.
            \begin{algorithm}[!t]
            \caption{Angular-Domain Nested Tensor Factor Estimation (NTFE-I)}

                \label{alg:proposed_1}
                \begin{algorithmic}[1]
                    \Require{Tensor $\boldsymbol{\mathcal{G}}$, $\boldsymbol{S}'$, maximum number of iterations $i_{\text{max}}$, and convergence threshold $\delta$.}      
                    \State{Randomly initialize $\hat{\bar{\boldsymbol{T}}}_{\boldsymbol{G}}$, $\hat{\boldsymbol{A}}_{\text{sr}}$, and $\hat{\boldsymbol{B}}_{\text{tx}}$ at iteration $i = 0$.}
                    \While{$||e(i) - e(i-1)|| \geq \delta$ and $i < i_{\text{max}}$}
                        \State{Find the \ac{LS} estimate of $\hat{\bar{\boldsymbol{T}}}_{\boldsymbol{G}}$ as} 
                        \begin{align*}
                            \hat{\bar{\boldsymbol{T}}}_{\boldsymbol{G}} = \left[\hat{\boldsymbol{\mathcal{G}}}\right]_{(1)} \left[\left(\boldsymbol{B}_{\text{tx}} \diamond \boldsymbol{A}_{\text{sr}}\right)^{\text{T}}\right]^{\dagger}
                        \end{align*}
                        \State{Find the \ac{LS} estimate of $\hat{\boldsymbol{A}}_{\text{sr}}$ as}
                        \begin{align*}
                            \hat{\boldsymbol{A}}_{\text{sr}} = \left[\hat{\boldsymbol{\mathcal{G}}}\right]_{(2)} \left[\left(\boldsymbol{B}_{\text{tx}} \diamond \bar{\boldsymbol{T}}_{\boldsymbol{G}}\right)^{\text{T}}\right]^{\dagger}
                        \end{align*}
                        \State{Find the \ac{LS} estimate of $\hat{\boldsymbol{B}}_{\text{tx}}$ as}
                        \begin{align*}
                            \hat{\boldsymbol{B}}_{\text{tx}} = \left[\hat{\boldsymbol{\mathcal{G}}}\right]_{(3)} \left[\left(\boldsymbol{A}_{\text{sr}} \diamond \bar{\boldsymbol{T}}_{\boldsymbol{G}}\right)^{\text{T}}\right]^{\dagger}
                        \end{align*}
                        \State{Update $e(i) = ||\hat{\boldsymbol{\mathcal{G}}} - \hat{\boldsymbol{\mathcal{G}}}(i)||^{2}_{\text{F}}$}
                    \EndWhile
                    \State{Compute the SVDs}
                    \begin{align*}
                        \hspace{-0,1cm}\hat{\boldsymbol{A}}_{\text{sr}} &= \sum^{K}_{k = 1} \sigma_{k} \boldsymbol{u}_{k} \boldsymbol{v}^{\text{H}}_{k}, \quad
                        \hat{\boldsymbol{B}}_{\text{tx}} = \sum^{K}_{k = 1} \sigma'_{k} \boldsymbol{u}'_{k} \boldsymbol{v}^{'\text{H}}_{k}
                    \end{align*}
                    \State{Define the estimates of the steering vectors}
                    \begin{align*}
                        \hat{\boldsymbol{a}}_{\text{sr}}(\phi^{k}_{\text{sr}},\theta^{k}_{\text{sr}})  = \sqrt{\sigma_{k}} \boldsymbol{u}_{k}, \hspace{0.25cm} \hat{\boldsymbol{b}}(\phi^{k}_{\text{ris}_{\text{D}}}, \theta^{k}_{\text{ris}_{\text{D}}}) = \sqrt{\sigma'_{k}} \boldsymbol{u}'_{k}
                    \end{align*}
                    \State{Apply $2$D \ac{ESPRIT} to $\hat{\boldsymbol{a}}_{\text{sr}}(\phi^{k}_{\text{sr}},\theta^{k}_{\text{sr}})$ and $\hat{\boldsymbol{b}}(\phi^{k}_{\text{ris}_{\text{A}}}, \theta^{k}_{\text{ris}_{\text{A}}})$ to estimate the angles at the \ac{SR} and the \ac{BD-RIS}. Refine the initial estimates}
                    \begin{align*}
                        \hat{\boldsymbol{A}}_{\text{sr}} &= \left[\boldsymbol{a}_{\text{sr}}(\hat{\phi}^{1}_{\text{sr}},\hat{\theta}^{1}_{\text{sr}}), \cdots, \boldsymbol{a}_{\text{sr}}(\hat{\phi}^{K}_{\text{sr}},\hat{\theta}^{K}_{\text{sr}})\right] \in \mathbb{C}^{L_{\text{SR}} \times K} \\
                        \hat{\boldsymbol{B}}_{\text{tx}} &= \left[\boldsymbol{b}_{\text{tx}}(\hat{\phi}^{1}_{\text{ris}_{\text{D}}},\hat{\theta}^{1}_{\text{ris}_{\text{D}}}), \cdots, \boldsymbol{b}_{\text{tx}}(\hat{\phi}^{K}_{\text{ris}_{\text{D}}},\hat{\theta}^{K}_{\text{ris}_{\text{D}}})\right] \in \mathbb{C}^{N \times K}
                    \end{align*}
                    \State{\textbf{return} $\hat{\boldsymbol{A}}_{\text{sr}}$, $ \hat{\boldsymbol{B}}_{\text{tx}}$, $\hat{\boldsymbol{\phi}}_{\text{sr}} = \left[\hat{\phi}^{1}_{\text{sr}}, \cdots, \hat{\phi}^{K}_{\text{sr}}\right]$, $\hat{\boldsymbol{\theta}}_{\text{sr}} = \left[\hat{\theta}^{1}_{\text{sr}}, \cdots, \hat{\theta}^{K}_{\text{sr}}\right]$, $\hat{\boldsymbol{\phi}}_{\text{ris}_{\text{D}}} = \left[\hat{\phi}^{1}_{\text{ris}_{\text{D}}}, \cdots, \hat{\phi}^{K}_{\text{ris}_{\text{D}}}\right]$, and  $\hat{\boldsymbol{\theta}}_{\text{ris}_{\text{D}}} = \left[\hat{\theta}^{1}_{\text{ris}_{\text{D}}}, \cdots, \hat{\theta}^{K}_{\text{ris}_{\text{D}}}\right]$.}
                \end{algorithmic}
            \end{algorithm} 
            
        \subsection{Decoupled Delay and Doppler Information Estimation}
            To estimate the nested model in (\ref{eq:tensor_J_1}) and the inner PARAFAC in (\ref{eq:tensor_F_1}), we solve the fitting problem
            \begin{equation}
                \hspace{-0.15cm}\left\{\hat{\boldsymbol{H}},[\hat{\boldsymbol{\mathcal{F}}}]_{(1)}\right\} = \underset{\boldsymbol{H},\left[\boldsymbol{\mathcal{F}}\right]_{(1)}}{\text{arg min}} \hspace{-0.075cm} \left| \hspace{-0.05cm} \left| \begin{split}&\hat{\boldsymbol{\mathcal{J}}} - \boldsymbol{\mathcal{I}}_{3,M Q} \times_{1} \boldsymbol{I}_{M Q} \\ &\times_{2} \boldsymbol{H} \boldsymbol{X} \times_{3} \left[\boldsymbol{\mathcal{F}}\right]_{(1)} \end{split} \right| \hspace{-0.05cm}\right|^{2}_{\text{F}},\label{eq:tensor_J_fit_problem_1}
            \end{equation}
            by employing the \ac{BALS} procedure, solving the following intermediate steps until a stoping criterion is achieved 
            \begin{align}
                 \hat{\boldsymbol{H}} &\hspace{-0.075cm}=\hspace{-0.075cm} \underset{\boldsymbol{H}}{\text{arg min}} \left|\left|[\hat{\boldsymbol{\mathcal{J}}}]_{(2)} \hspace{-0.075cm}-\hspace{-0.075cm} \boldsymbol{H} \boldsymbol{X} \left(\left[\boldsymbol{\mathcal{F}}\right]_{(1)} \hspace{-0.075cm}\diamond\hspace{-0.075cm} \boldsymbol{I}_{M Q}\right)^{\text{T}}\right|\right|^{2}_{\text{F}}, \label{eq:tensor_J_1_ls_1} \\
                 &= \left[\boldsymbol{\mathcal{J}}\right]_{(2)} \left(\boldsymbol{X} \left(\left[\boldsymbol{\mathcal{F}}\right]_{(1)} \diamond \boldsymbol{I}_{M Q}\right)^{\text{T}}\right)^{\dagger}, \label{eq:tensor_J_1_pseudoinverse_1} \\
                 [\hat{\boldsymbol{\mathcal{J}}}]_{(1)} &\hspace{-0.075cm}=\hspace{-0.075cm} \underset{\left[\boldsymbol{\mathcal{F}}\right]_{(1)}}{\text{arg min}} \left|\left|[\hat{\boldsymbol{\mathcal{J}}}]_{(3)} \hspace{-0.075cm}-\hspace{-0.075cm} \left[\boldsymbol{\mathcal{F}}\right]_{(1)} \left( \boldsymbol{H} \boldsymbol{X} \hspace{-0.075cm}\diamond\hspace{-0.075cm} \boldsymbol{I}_{M Q}\right)^{\text{T}}\right|\right|^{2}_{\text{F}}, \label{eq:tensor_J_1_ls_2} \\
                 &= \left[\boldsymbol{\mathcal{J}}\right]_{(3)} \left( \left(\boldsymbol{H} \boldsymbol{X}  \diamond \boldsymbol{I}_{M Q}\right)^{\text{T}}\right)^{\dagger}, \label{eq:tensor_J_2_pseudoinverse_1}
            \end{align}
            The above solutions are ensured by the condition $M Q\geq L_{\text{ST}}$. 
            
            Note that we assume knowledge of the departure angles at the \ac{ST}, $\phi_{\text{st}}$ and $\theta_{\text{st}}$, otherwise identifiability arise in the second processing stage. This is discussed in detail in a later section. Thus, from the \ac{ST}-\ac{BD-RIS} channel estimate we can define $\hat{\boldsymbol{H}} = \sigma \boldsymbol{u} \boldsymbol{v}^{\text{H}}$, where we employ $2$D \ac{ESPRIT} on $\hat{\boldsymbol{b}}_{\text{rx}}(\phi_{\text{ris}_{\text{A}}}, \theta_{\text{ris}_{\text{A}}}) = \sqrt{\sigma} \boldsymbol{u}$ to obtain the estimates of the arrival angles at the \ac{BD-RIS}, $\hat{\phi}_{\text{ris}_{\text{A}}}$ and $\hat{\theta}_{\text{ris}_{\text{A}}}$. 
            
            Acquisition of the targets' delay and Doppler requires exploiting the structure of the PARAFAC tensor defined in (\ref{eq:tensor_F_1}). First, we explore the following fitting problem:
            \begin{equation}
                \hspace{-0.15cm}\left\{\hat{\boldsymbol{T}}_{\boldsymbol{J}}, \hat{\boldsymbol{D}}_{\nu}, \boldsymbol{C}_{\tau}\right\} = \underset{\hat{\boldsymbol{T}}_{\boldsymbol{J}}, \boldsymbol{D}_{\nu}, \boldsymbol{C}_{\tau}}{\text{arg min}} \hspace{-0.075cm} \left| \hspace{-0.05cm} \left| \begin{split}&\hat{\boldsymbol{\mathcal{F}}} -  \boldsymbol{\mathcal{I}}_{3,K} \times_{1} \boldsymbol{T}_{\boldsymbol{J}} \\&\times_{2} \boldsymbol{D}_{\nu} \times_{3} \boldsymbol{C}_{\tau}\end{split} \right| \hspace{-0.05cm}\right|^{2}_{\text{F}},\label{eq:tensor_F_fit_problem_1}
            \end{equation}
            which we solve using \ac{ALS} applied sequentially to the individual estimates 
            \begin{align}
                \hat{\boldsymbol{T}}_{\boldsymbol{J}} &\hspace{-0.075cm}=\hspace{-0.075cm} \underset{\boldsymbol{D}_{\nu}}{\text{arg min}} \left|\left|[\hat{\boldsymbol{\mathcal{J}}}]_{(1)} - \boldsymbol{T}_{\boldsymbol{J}} \left(\boldsymbol{C}_{\tau} \diamond \boldsymbol{D}_{\nu}\right)^{\text{T}} \right|\right|^{2}_{\text{F}}, \label{eq:tensor_F_1_ls_1} \\
                &= [\hat{\boldsymbol{\mathcal{J}}}]_{(1)} \left(\left(\boldsymbol{C}_{\tau} \diamond \boldsymbol{D}_{\nu}\right)^{\text{T}}\right)^{\dagger}, \label{eq:tensor_F_1_pseudoinverse_1} \\
                \hat{\boldsymbol{D}}_{\nu} &\hspace{-0.075cm}=\hspace{-0.075cm} \underset{\boldsymbol{D}_{\nu}}{\text{arg min}} \left|\left|[\hat{\boldsymbol{\mathcal{J}}}]_{(2)} - \boldsymbol{D}_{\nu} \left(\boldsymbol{C}_{\tau} \diamond \boldsymbol{T}_{\boldsymbol{J}}\right)^{\text{T}} \right|\right|^{2}_{\text{F}}, \label{eq:tensor_F_1_ls_2} \\
                &= [\hat{\boldsymbol{\mathcal{J}}}]_{(2)} \left(\left(\boldsymbol{C}_{\tau} \diamond \boldsymbol{T}_{\boldsymbol{J}}\right)^{\text{T}}\right)^{\dagger}, \label{eq:tensor_F_1_pseudoinverse_2} \\
                 \hat{\boldsymbol{C}}_{\tau} &\hspace{-0.075cm}=\hspace{-0.075cm} \underset{\boldsymbol{C}_{\tau}}{\text{arg min}} \left|\left|[\hat{\boldsymbol{\mathcal{J}}}]_{(3)} - \boldsymbol{C}_{\tau} \left(\boldsymbol{D}_{\nu} \diamond \boldsymbol{T}_{\boldsymbol{J}}\right)^{\text{T}} \right|\right|^{2}_{\text{F}}, \label{eq:tensor_F_1_ls_3} \\
                 &= [\hat{\boldsymbol{\mathcal{J}}}]_{(3)} \left(\left(\boldsymbol{D}_{\nu} \diamond \boldsymbol{T}_{\boldsymbol{J}}\right)^{\text{T}}\right)^{\dagger}. \label{eq:tensor_F_1_pseudoinverse_3}
            \end{align}
            The existence of the above solutions is ensured by the condition $M Q \geq K$ condition. After obtaining the estimates $\hat{\boldsymbol{D}}_{\nu}$ and $\hat{\boldsymbol{C}}_{\tau}$, we apply $1$D \ac{ESPRIT} to acquire the delay and Doppler estimates for the $K$ targets. The summary of the proposed solution is given in Algorithm \ref{alg:proposed_2}.      
            \begin{algorithm}[!t]
                \caption{Delay-Doppler Nested Tensor Factor Estimation (NTFE-II)}  
                \label{alg:proposed_2}
                \begin{algorithmic}[1]
                    \Require{Tensor $\boldsymbol{\mathcal{J}}$, $\boldsymbol{X}$, maximum number of iterations $i_{\text{max}}$, and convergence threshold $\delta$.}    
                    \State{Randomly initialize $\hat{\boldsymbol{H}}$ and $[\hat{\boldsymbol{\mathcal{J}}}]_{(1)}$ at iteration $i = 0$.}
                    \While{$||e(i) - e(i-1)|| \geq \delta$ and $i < i_{\text{max}}$}
                        \State{Calculate the \ac{LS} estimate}
                        \begin{align*}
                            \hat{\boldsymbol{H}} = \left[\boldsymbol{\mathcal{J}}\right]_{(2)} \left(\boldsymbol{X} \left(\left[\boldsymbol{\mathcal{F}}\right]_{(1)} \diamond \boldsymbol{I}_{M Q}\right)^{\text{T}}\right)^{\dagger}
                        \end{align*}
                        \State{Calculate the \ac{LS} estimate}
                        \begin{align*}
                            [\hat{\boldsymbol{\mathcal{J}}}]_{(1)} = \left[\boldsymbol{\mathcal{J}}\right]_{(3)} \left( \left(\boldsymbol{H} \boldsymbol{X}  \diamond \boldsymbol{I}_{M Q}\right)^{\text{T}}\right)^{\dagger}
                        \end{align*}
                        \State{Update $e(i) = ||\hat{\boldsymbol{\mathcal{Y}}} - \hat{\boldsymbol{\mathcal{Y}}}(i)||^{2}_{\text{F}}$}
                    \EndWhile
                    \State{Compute the SVD}
                    \begin{align*}
                        \hat{\boldsymbol{H}} = \sigma \boldsymbol{u} \boldsymbol{v}^{\text{H}}
                    \end{align*}
                    \State{Define the steering vector estimate at the \ac{BD-RIS}}
                    \begin{align*}
                        \hat{\boldsymbol{b}}_{\text{rx}}(\phi_{\text{ris}_{\text{A}}}, \theta_{\text{ris}_{\text{A}}}) = \sqrt{\sigma} \boldsymbol{u}
                    \end{align*}
                    \State{Apply $2$D \ac{ESPRIT} to $\hat{\boldsymbol{b}}_{\text{rx}}(\phi_{\text{ris}_{\text{A}}}, \theta_{\text{ris}_{\text{A}}})$ to acquire $\hat{\phi}_{\text{ris}_{\text{A}}}$ and $\hat{\theta}_{\text{ris}_{\text{A}}}$. Define tensor $\hat{\boldsymbol{\mathcal{F}}}$ from (\ref{eq:tensor_J_1})}.
                    \State{Randomly initialize $\hat{\boldsymbol{T}}_{\boldsymbol{J}}$, $\hat{\boldsymbol{D}}_{\nu}$, and $\hat{\boldsymbol{C}}_{\tau}$ at iteration $i = 0$.}
                    \While{$||e(i) - e(i-1)|| \geq \delta$ and $i < i_{\text{max}}$}
                        \State{Calculate the \ac{LS} estimate}
                        \begin{align*}
                            \hat{\boldsymbol{T}}_{\boldsymbol{J}} = [\hat{\boldsymbol{\mathcal{J}}}]_{(1)} \left(\left(\boldsymbol{C}_{\tau} \diamond \boldsymbol{D}_{\nu}\right)^{\text{T}}\right)^{\dagger}
                        \end{align*}
                        \State{Calculate the \ac{LS} estimate}
                        \begin{align*}
                            \hat{\boldsymbol{D}}_{\nu} = [\hat{\boldsymbol{\mathcal{J}}}]_{(2)} \left(\left(\boldsymbol{C}_{\tau} \diamond \boldsymbol{T}_{\boldsymbol{J}}\right)^{\text{T}}\right)^{\dagger}
                        \end{align*}
                         \State{Calculate the \ac{LS} estimate}
                         \begin{align*}
                            \hat{\boldsymbol{C}}_{\tau} = [\hat{\boldsymbol{\mathcal{J}}}]_{(3)} \left(\left(\boldsymbol{D}_{\nu} \diamond \boldsymbol{T}_{\boldsymbol{J}}\right)^{\text{T}}\right)^{\dagger}
                         \end{align*}
                        \State{Update $e(i) = ||\hat{\boldsymbol{\mathcal{Y}}} - \hat{\boldsymbol{\mathcal{Y}}}(i)||^{2}_{\text{F}}$}
                    \EndWhile
                    \State{Obtain $\hat{\boldsymbol{\tau}} = \left[\hat{\tau}_{1}, \cdots, \hat{\tau}_{K}\right]$ and $\hat{\boldsymbol{\nu}} = \left[\hat{\nu}_{1}, \cdots, \hat{\nu}_{K}\right]$ from $\hat{\boldsymbol{C}}(\tau)$ and $\hat{\boldsymbol{D}}(\nu)$ using $1$D \ac{ESPRIT}.}
                    \State{Update the estimates of $\boldsymbol{H}$, $\boldsymbol{D}_{\nu}$, and $\boldsymbol{C}_{\tau}$ as}
                    \begin{align*}
                        \hat{\boldsymbol{H}} &= \boldsymbol{b}_{\text{rx}}(\hat{\phi}_{\text{ris}_{\text{A}}},\hat{\theta}_{\text{ris}_{\text{A}}}) \boldsymbol{a}^{\text{T}}_{\text{st}}(\phi_{\text{st}},\hspace{-0.05cm}\theta_{\text{st}}) \\
                        \hat{\boldsymbol{D}}_{\nu} &= \left[\boldsymbol{d}(\hat{\nu}_{1}), \cdots, \boldsymbol{d}(\hat{\nu}_{K})\right] \\
                        \hat{\boldsymbol{C}}_{\tau} &= \left[\boldsymbol{c}(\hat{\tau}_{1}), \cdots, \boldsymbol{c}(\hat{\tau}_{K})\right]
                    \end{align*}
                    \State{\textbf{return} $\hat{\boldsymbol{H}}$, $\hat{\boldsymbol{D}}_{\nu}$, $\hat{\boldsymbol{C}}_{\tau}$, $\hat{\boldsymbol{\tau}}$, $\hat{\boldsymbol{\nu}}$, $\hat{\phi}_{\text{ris}_{\text{A}}}$, and $\hat{\theta}_{\text{ris}_{\text{A}}}$.}
                \end{algorithmic}
            \end{algorithm}
        \subsection{Closed-form Complex Channel Coefficient Estimation}
\begin{figure}[!t]
                \centering
                \includegraphics[width=0.985\columnwidth]{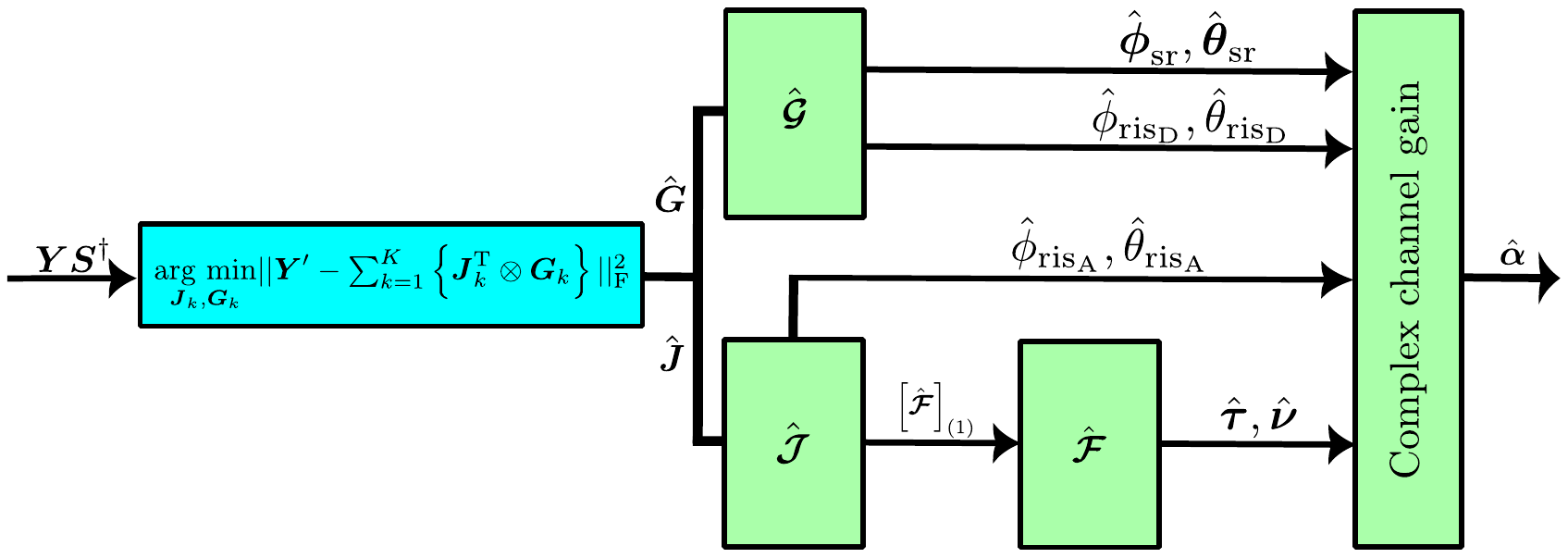}
                \caption{Block diagram of the proposed \ac{TenDAE} framework: (i) \acf{KSA}, (ii) \acf{NTFE} for joint estimation of the tensors $\boldsymbol{G}$ and $\boldsymbol{J}$, and (iii) parameter extraction using, e.g., \ac{ESPRIT}. First, we solve the \ac{KSA} problem (\ref{eq:ksa_1}) after right-filtering the received signal with the \ac{BD-RIS} phase-shift matrix. Then, we estimate the tensors in (\ref{eq:tensor_G_1}) and (\ref{eq:tensor_J_1}) with Algorithms \ref{alg:proposed_1} and \ref{alg:proposed_2}, respectively. Finally, we leverage the estimated parameters to extract the complex channel coefficients using (\ref{eq:complex channel coefficient_estimation}).}
                \label{fig:solution_diagram}
            \end{figure}

            Given the angle, delay, and Doppler estimates, we finally recover the complex gains $\alpha_{k}$. From (\ref{eq:received_signal_filtered}), we form the reconstructed signal
            \begin{align}
                \hat{\boldsymbol{Y}}' = \sum^{K}_{k = 1} \{\hat{\boldsymbol{J}}^{T}_{k} \otimes \hat{\boldsymbol{G}}_{k}\},
            \end{align}
            which can be recast using the rearrangement operator as
            \begin{align}
                \mathcal{R}_{M Q \times N}(\hat{\boldsymbol{Y}}') &= \hat{\boldsymbol{G}} \hat{\boldsymbol{J}}^{T} = \bar{\boldsymbol{G}} \boldsymbol{\Lambda}_{\boldsymbol{\alpha}} \hat{\boldsymbol{T}}_{\boldsymbol{G}} \hat{\boldsymbol{T}}^{\text{T}}_{\boldsymbol{J}} \bar{\boldsymbol{J}} =  \bar{\boldsymbol{G}} \Lambda_{\alpha} \bar{\boldsymbol{J}}, \label{eq:complex channel coefficient_estimation_equation}
            \end{align}
            where $\bar{\boldsymbol{G}} = (\hat{\boldsymbol{B}}_{\text{tx}} \diamond \hat{\boldsymbol{A}}_{\text{sr}}) \in \mathbb{C}^{L_{\text{SR}} N \times K}$ and $\bar{\boldsymbol{J}} = \left(\hat{\boldsymbol{C}}_{\tau} \diamond \hat{\boldsymbol{D}}_{\nu}\right)^{\text{T}} \left((\hat{\boldsymbol{H}} \boldsymbol{X}) \diamond \boldsymbol{I}_{M Q}\right)^{\text{T}} \in \mathbb{C}^{K \times M Q N}$ are the refined \ac{CSI} estimates after acquisition of the target parameters. The transformation matrices from the initial \ac{KSA} problem satisfy $\hat{\boldsymbol{T}}_{\boldsymbol{G}} \hat{\boldsymbol{T}}^{\text{T}}_{\boldsymbol{J}} = \boldsymbol{I}_{K}$, and thus cancel each other. Initially, $\bar{\boldsymbol{G}}$ is reconstructed using the estimates from Alg.~\ref{alg:proposed_1} and $\bar{\boldsymbol{J}}$ is reconstructed from the estimates from Alg.~\ref{alg:proposed_2}. We then apply Property (\ref{eq:property_3}) to (\ref{eq:complex channel coefficient_estimation_equation}) to obtain 
            \begin{align}
                \text{vec}(\mathcal{R}_{M Q \times N}(\hat{\boldsymbol{Y}}')) &= ( \bar{\boldsymbol{J}}^{\text{T}} \diamond \bar{\boldsymbol{G}}) \boldsymbol{\alpha},
            \end{align}
            from which we get an estimate of $\boldsymbol{\alpha}$ by left filtering $\text{vec}(\mathcal{R}_{M Q \times N}(\hat{\boldsymbol{Y}}'))$ as
            \begin{align}
                \hat{\boldsymbol{\alpha}} &= ( \bar{\boldsymbol{J}}^{\text{T}} \diamond \bar{\boldsymbol{G}})^{\dagger} \text{vec}(\mathcal{R}_{M Q \times N}(\hat{\boldsymbol{Y}}')), \label{eq:complex channel coefficient_estimation}
            \end{align}
            ensured by the condition $L_{\text{SR}} N M Q N \geq K$. In Fig. \ref{fig:solution_diagram}, we present a simplified pictogram of the proposed methodology for delay-Doppler-angle estimation. The proposed \ac{TenDAE} framework is summarized in Alg.~\ref{alg:proposed_3}.
                \begin{algorithm}[!t]
                \caption{\Acf{TenDAE}}  
                \label{alg:proposed_3}
                \begin{algorithmic}[1]
                    \Require{Tensor $\boldsymbol{\mathcal{G}}$, $\boldsymbol{\mathcal{J}}$, $\boldsymbol{S}'$, $\boldsymbol{X}$, maximum number of iterations $i_{\text{max}}$, and convergence threshold $\delta$.}    
                    \State{Right-filter the phase-shift matrix $\boldsymbol{S}'$}
                    \begin{align*}
                        \boldsymbol{Y}' = \boldsymbol{Y} \boldsymbol{S}^{\dagger}
                    \end{align*}
                    \State{Find the solution of the following \ac{KSA} problem}
                    \begin{align*}
                        \{\hat{\boldsymbol{J}}_{k}, \hat{\boldsymbol{G}}_{k}\} = \underset{\boldsymbol{J}_{k}, \boldsymbol{G}_{k}}{\text{arg min}} \left|\left|\begin{aligned}\boldsymbol{Y}' -\sum^{K}_{k = 1} \left\{ \boldsymbol{J}^{\text{T}}_{k} \otimes \boldsymbol{G}_{k}\right\}\end{aligned}\right|\right|^{2}_{\text{F}}
                    \end{align*}
                    \State{Apply \ac{NTFE}-I to extract the $K$ target angles angular.}
                    \State{Apply \ac{NTFE}-II to extract the $K$ target delays/Dopplers.}
                    \State{Estimate the complex channel gain}
                    \begin{align*}
                        \hat{\boldsymbol{\alpha}} &= ( \bar{\boldsymbol{J}}^{\text{T}} \diamond \bar{\boldsymbol{G}})^{\dagger} \text{vec}(\mathcal{R}_{M Q \times N}(\hat{\boldsymbol{Y}}'))
                    \end{align*}
                    \State{\textbf{return} $\hat{\boldsymbol{\tau}}$, $\hat{\boldsymbol{\nu}}$, $\hat{\phi}_{\text{ris}_{\text{A}}}$, $\hat{\theta}_{\text{ris}_{\text{A}}}$, $\hat{\boldsymbol{\phi}}_{\text{sr}}$, $\hat{\boldsymbol{\theta}}_{\text{sr}}$, $\hat{\boldsymbol{\phi}}_{\text{ris}_{\text{D}}}$, and $\hat{\boldsymbol{\theta}}_{\text{ris}_{\text{D}}}$.}
                \end{algorithmic}
            \end{algorithm}

    \section{Identifiability and Uniqueness Conditions} \label{sec:identifiability_and_uniqueness_conditions}
        Here we summarize identifiability conditions that guarantee unique recovery of the target parameters from Alg.~\ref{alg:proposed_1} and Alg.~\ref{alg:proposed_2}. We begin by recalling that the rank-$K$ \ac{KSA} estimate in (\ref{eq:ksa_1}) is \ac{SVD}-based and therefore non-unique up to nonsingular transformations. In particular, for $\boldsymbol{T}_{\boldsymbol{G}},\boldsymbol{T}_{\boldsymbol{J}} \in \mathbb{C}^{K \times K}$,
        \begin{align}
            \mathcal{R}_{M Q \times N}(\boldsymbol{Y}') = \hat{\boldsymbol{G}} \hat{\boldsymbol{J}}^{\text{T}} = \boldsymbol{G} \boldsymbol{T}_{\boldsymbol{G}} \boldsymbol{T}_{\boldsymbol{J}} \boldsymbol{J}^{\text{T}} = \boldsymbol{G} \boldsymbol{J}^{\text{T}},
        \end{align}
        so any pair $(\hat{\boldsymbol{G}},\hat{\boldsymbol{J}})$ related by compensating transforms yields the same reconstruction.

        For the PARAFAC model in (\ref{eq:tensor_G_1}), the factors are identifiable up to diagonal scalings, i.e., $\hat{\bar{\boldsymbol{T}}}_{\text{sr}} = \bar{\boldsymbol{T}}_{\boldsymbol{G}} \boldsymbol{\Lambda}_{1}$, $\hat{\boldsymbol{A}}_{\text{sr}} = \boldsymbol{A}_{\text{sr}} \boldsymbol{\Lambda}_{2}$, and $\hat{\boldsymbol{B}}_{\text{tx}} = \boldsymbol{B}_{\text{tx}} \boldsymbol{\Lambda}_{3}$, with $\boldsymbol{\Lambda}_{1} \boldsymbol{\Lambda}_{2} \boldsymbol{\Lambda}_{3} = \boldsymbol{I}_{K}$. Similarly, for (\ref{eq:tensor_J_1}) and (\ref{eq:tensor_F_1}), we have $\hat{\boldsymbol{H}} = \boldsymbol{H} \boldsymbol{\Lambda}_{4}$, $[\hat{\boldsymbol{\mathcal{F}}}]_{(1)} = \left[\boldsymbol{\mathcal{F}}\right]_{(1)} \boldsymbol{\Lambda}_{5}$ with $\boldsymbol{\Lambda}_{4} \boldsymbol{\Lambda}_{5} = \boldsymbol{I}_{K}$, and $\hat{\boldsymbol{T}}_{\boldsymbol{J}} = \boldsymbol{T}_{\boldsymbol{J}} \boldsymbol{\Lambda}_{6}$, $\hat{\boldsymbol{D}}_{\nu} = \boldsymbol{D}_{\nu} \boldsymbol{\Lambda}_{7}$, $\hat{\boldsymbol{C}}_{\tau} = \boldsymbol{C}_{\tau} \boldsymbol{\Lambda}_{8}$ with $\boldsymbol{\Lambda}_{6} \boldsymbol{\Lambda}_{7} \boldsymbol{\Lambda}_{8} = \boldsymbol{I}_{K}$. As discussed in Section \ref{sec:tensor_based_multi_target_sensing}, identifiability hinges on the right-invertibility of the linear systems that appear in the \ac{LS} updates of Alg. \ref{alg:proposed_1} and Alg. \ref{alg:proposed_2}. For Alg. \ref{alg:proposed_1}, define
        $\boldsymbol{P}_{1} = \left(\boldsymbol{B}_{\text{tx}} \diamond \boldsymbol{A}_{\text{sr}}\right)^{\text{T}} \in \mathbb{C}^{K \times L_{\text{SR}} N}$,
        $\boldsymbol{P}_{2} = \left(\boldsymbol{A}_{\text{sr}} \diamond \bar{\boldsymbol{T}}_{\boldsymbol{G}} \right)^{\text{T}} \in \mathbb{C}^{K \times K L_{\text{SR}}}$, and
        $\boldsymbol{P}_{3} = \left(\boldsymbol{B}_{\text{tx}} \diamond \bar{\boldsymbol{T}}_{\boldsymbol{G}}\right)^{\text{T}} \in \mathbb{C}^{K \times K N}$.
        Then the \ac{LS} updates in (\ref{eq:tensor_G_1_ls_1})--(\ref{eq:tensor_G_1_ls_3}) require $\boldsymbol{P}_{1}$, $\boldsymbol{P}_{2}$, and $\boldsymbol{P}_{3}$ to be right-invertible, leading to $L_{\text{SR}} N \geq K$ (with $L_{\text{SR}}\geq 1$ and $N\geq 1$).

        For Alg. \ref{alg:proposed_2}, the \ac{BALS} stage uses (\ref{eq:tensor_J_1_ls_1})--(\ref{eq:tensor_J_1_ls_2}). Defining $\boldsymbol{P}_{4} = \boldsymbol{X} \left(\left[\boldsymbol{\mathcal{F}}\right]_{(1)} \diamond \boldsymbol{I}_{M Q}\right)^{\text{T}} \in \mathbb{C}^{L_{\text{ST}} \times M Q}$ and $\boldsymbol{P}_{5} = \left(\boldsymbol{H} \boldsymbol{X}  \diamond \boldsymbol{I}_{M Q}\right)^{\text{T}} \in \mathbb{C}^{M Q \times M Q N}$, identifiability requires $M Q \geq L_{\text{ST}}$ and $N \geq 1$. In the subsequent \ac{ALS} stage, define $\boldsymbol{P}_{6} = \left(\boldsymbol{C}_{\tau} \diamond \boldsymbol{D}_{\nu}\right)^{\text{T}} \in \mathbb{C}^{K \times M Q}$, $\boldsymbol{P}_{7} = \left(\boldsymbol{C}_{\tau} \diamond \boldsymbol{T}_{\boldsymbol{J}}\right)^{\text{T}}\in \mathbb{C}^{K \times Q}$, and $\boldsymbol{P}_{8} = \left(\boldsymbol{D}_{\nu} \diamond \boldsymbol{T}_{\boldsymbol{J}}\right)^{\text{T}}\in \mathbb{C}^{K \times M}$. Right-invertibility yields $M Q \geq K$ (with $Q\geq 1$ and $M\geq 1$). Finally, for \ac{LS} gain estimation in (\ref{eq:complex channel coefficient_estimation}), let $\boldsymbol{P}_{9} = ( \bar{\boldsymbol{J}}^{\text{T}} \diamond \bar{\boldsymbol{G}}) \in \mathbb{C}^{L_{\text{SR}} N M Q N \times K}$. Left-invertibility requires $L_{\text{SR}} N M Q N \geq K$. Overall, identifiability is ensured by $M Q \geq \max(L_{\text{ST}}, K)$ and $L_{\text{SR}} N M Q N \geq K$.
        
        \subsubsection*{Sufficient condition} Since our approach relies on three PARAFAC decompositions, we can invoke Kruskal's uniqueness condition \cite{kruskal1977three,kruskal1989rank}. For an $N$th-order tensor $\boldsymbol{\mathcal{X}} = \boldsymbol{\mathcal{I}}_{N,R} \times_{1} \boldsymbol{A}_{1} \times_{2} \cdots \times_{N} \boldsymbol{A}_{N}$, uniqueness is guaranteed if
            $\sum^{N}_{n = 1} k_{\boldsymbol{A}_{n}} \geq 2 R + (N - 1)$,
        where $k_{\boldsymbol{A}_{n}}$ is the $k$-rank\footnote{The $k$-rank is the maximum value $k$ such that any $k$ columns are linearly independent \cite{kruskal1977three}.} of $\boldsymbol{A}_{n} \in \mathbb{C}^{I_{n} \times R}$. Applying this condition to each decomposition (\ref{eq:tensor_G_1}), (\ref{eq:tensor_J_1}), and (\ref{eq:tensor_F_1}), the conditions ensuring uniqueness are $K \geq 2$, $K \geq M Q + 1$, and $K \geq 2$, respectively.
  
        \begin{table*}[!t]
            \centering
            \caption{Estimation Computational Complexity}
            \label{tab:complexity}
            \resizebox{1\textwidth}{!}{
            \begin{tabular}{|c|c|}
            \hline
            \textbf{Algorithm}                           & \textbf{Computational Complexity} \\ \hline
            \acs{LS} (\ref{eq:received_signal_filtered}) & $\mathcal{O}(N^{4} T)$            \\ \hline
            \acs{KSA} (\ref{eq:ksa_1})                   & $\mathcal{O}(L_{\text{SR}} M Q N^{2} K)$  \\ \hline
            \ac{NTFE} - I (HOSVD \cite{benicio2024low})                   & $\mathcal{O}(L_{\text{SR}} N K(L_\text{SR} + N + K))$       \\ \hline
            \ac{NTFE} - II (HOSVD \cite{benicio2024low})                  & $\mathcal{O}(M Q N K(M Q + N + K) + 3 M Q K^2)$ \\ \hline
            \ac{NTFE} - I (ALS)                   & $\mathcal{O}((\frac{L^{2}_{\text{SR}}N}{K} + L_{\text{SR}} + N)K^{2} \text{ALS}_{\text{iter}} + 2(L_{\text{SR}} + N)K + (5L_{\text{SR}} + 3N + 1)K^2 + 6K^{3})$       \\ \hline
            \ac{NTFE} - II (ALS)                  & $\mathcal{O}([(M Q N+ K^{2} L_{\text{ST}})M Q + (\frac{MQ}{K} + M + Q)K^{3}]\text{ALS}_{\text{iter}} + (3 + 2L_{\text{ST}})N + (M + N + Q + 1)K + 3(M + Q)K^{2} + 6 K^{3})$             
            \\ \hline
            \end{tabular}
            }
        \end{table*}

    \section{Computational Complexity} \label{sec:Computational Complexity}

        Table \ref{tab:complexity} summarizes the dominant computational costs of the considered channel estimation methods (\ac{LS}, \ac{KSA}, and \ac{KRSA}), and the proposed Algorithms~\ref{alg:proposed_1}--\ref{alg:proposed_2} \textit{via} either \ac{ALS} or \ac{HOSVD}.
        We assume a complexity of order $\mathcal{O}(m n^{2})$ for the pseudoinverse of $\mathbf{A}\in\mathbb{C}^{m\times n}$ ($m>n$) and $\mathcal{O}(m n r)$ for a rank-$r$ \ac{SVD} approximation, omitting constant factors. In our setting, \ac{LS} estimates $\boldsymbol{H}_{\text{eff}} = \boldsymbol{Y} \boldsymbol{S}^{\dagger}$, whereas \ac{KSA} estimates $\boldsymbol{H}_{\text{eff}} = \sum^{K}_{k = 1} \left\{\boldsymbol{J}^{\text{T}}_{k} \otimes \boldsymbol{G}_{k}\right\}$ via a rank-$K$ \ac{SVD}. Thus, the complexity of \ac{LS}, \ac{KSA}, and \ac{KRSA} scale as $\mathcal{O}(N^{4} T)$, $\mathcal{O}(L_{\text{SR}} M Q N^{2} K)$, and $\mathcal{O}(L_{\text{SR}} M Q N K)$, respectively.
        \subsection{Complexity of NTFE-I (angle estimation)}
            Algorithm \ref{alg:proposed_1} consists of three stages: (i) initialization, (ii) \ac{ALS} updates for the PARAFAC fit, and (iii) parameter extraction via \ac{SVD}/\ac{ESPRIT}. Random initialization of $\hat{\bar{\boldsymbol{T}}}_{\boldsymbol{G}}$, $\hat{\boldsymbol{A}}_{\text{sr}}$, and $\hat{\boldsymbol{B}}_{\text{tx}}$ requires complexity of order $\mathcal{O}((L_{\text{SR}}+N+K)K)$. The \ac{ALS} loop evaluates the three pseudoinverses in (\ref{eq:tensor_G_1_pseudoinverse_1})--(\ref{eq:tensor_G_1_pseudoinverse_3}) over $\text{ALS}_{\text{iter}}$ iterations, with per-iteration cost $\mathcal{O}\left(\left(\frac{L_{\text{SR}} N}{K} + N + L_{\text{SR}}\right)K^{2}\right)$. The extraction stage applies two rank-$K$ \ac{SVD} with cost $\mathcal{O}((L_{\text{SR}}+N)K^{2})$ and two $2$D \ac{ESPRIT} procedures with costs $\mathcal{O}\left(\left(\frac{N}{K^{2}} + \frac{2N}{K} + 3\right)K^{3}\right)$ (for $\hat{\boldsymbol{B}}_{\text{tx}}$) and $\mathcal{O}\left(\left(\frac{L_{\text{SR}}}{K^{2}} + \frac{2L_{\text{SR}}}{K} + 3\right)K^{3}\right)$ (for $\hat{\boldsymbol{A}}_{\text{sr}}$). Therefore, the overall complexity of Alg.~\ref{alg:proposed_1} is $\mathcal{O}((\frac{L^{2}_{\text{SR}}N}{K} + L_{\text{SR}} + N)K^{2} \text{ALS}_{\text{iter}} + 2(L_{\text{SR}} + N)K + (5L_{\text{SR}} + 3N + 1)K^2 + 6K^{3})$. 
        \subsection{Complexity of NTFE-II (delay-Doppler estimation)}
            Algorithm \ref{alg:proposed_2} comprises two iterative stages, \ac{BALS} for $\hat{\boldsymbol{H}}$ and $[\hat{\boldsymbol{\mathcal{J}}}]_{(1)}$, followed by \ac{ALS} for $\hat{\boldsymbol{T}}_{\boldsymbol{J}}$, $\hat{\boldsymbol{D}}_{\nu}$, and $\hat{\boldsymbol{C}}_{\tau}$, and a final parameter-extraction step. In the \ac{BALS} stage, the random initialization is of order $\mathcal{O}(N L_{\text{ST}} + M Q K)$, and the two pseudoinverses in (\ref{eq:tensor_J_1_ls_1})--(\ref{eq:tensor_J_1_ls_2}) require $\mathcal{O}((M Q N+ K^{2} L_{\text{ST}})M Q)$ operations per iteration over $\text{ALS}_{\text{iter}}$ iterations. The complexity of the subsequent rank-one \ac{SVD} of $\hat{\boldsymbol{H}}$ is $\mathcal{O}(N L_{\text{ST}})$ and $\mathcal{O}(3N)$ for the associated $2$D \ac{ESPRIT}. In the \ac{ALS} stage, initializing $\hat{\boldsymbol{T}}_{\boldsymbol{J}}$, $\hat{\boldsymbol{D}}_{\nu}$, and $\hat{\boldsymbol{C}}_{\tau}$ requires $\mathcal{O}((K + N + M)K)$ operations. The three pseudoinverses in (\ref{eq:tensor_F_1_pseudoinverse_1})--(\ref{eq:tensor_F_1_pseudoinverse_3}) have a per-iteration complexity of $\mathcal{O}((\frac{MQ}{K} + M + Q)K^{3})$. Finally, delay and Doppler are extracted via a rank-$K$ \ac{SVD} followed by $1$D \ac{ESPRIT}, with complexity $\mathcal{O}(MK^{2}) + \mathcal{O}(2MK^{2} + 3K^{3})$ for Doppler and $\mathcal{O}(QK^{2}) + \mathcal{O}(2QK^{2} + 3K^{3})$ for delay. Therefore, the total complexity of Alg. \ref{alg:proposed_2} is $\mathcal{O}([(M Q N+ K^{2} L_{\text{ST}})M Q + (\frac{MQ}{K} + M + Q)K^{3}]\text{ALS}_{\text{iter}} + (3 + 2L_{\text{ST}})N + (M + N + Q + 1)K + 3(M + Q)K^{2} + 6 K^{3})$. The closed-form gain estimation in (\ref{eq:complex channel coefficient_estimation}) requires an additional pseudoinverse $(\bar{\boldsymbol{J}}^{\text{T}} \diamond \bar{\boldsymbol{G}})^{\dagger}$ with complexity $\mathcal{O}(L_{\text{SR}} M Q N^{2} K)$.\footnote{The parameter estimation usually takes between 2 to 10 iterations to converge depending on the noise level. Since we have a nested-PARAFAC in stage two, 10 iterations may be required for each tensor estimation problem for the worst noise level considered in this paper.}

    \section{Cramér-Rao Lower Bound} \label{sec:crlb}

        The model in (\ref{eq:received_signal_all_samples}) follows a multivariate complex Gaussian 
        \ac{PDF} parameterized by the vector
        \begin{align}
            \boldsymbol{\eta} &= \left[\boldsymbol{\alpha}_{\text{Re}}, \boldsymbol{\alpha}_{\text{Im}}, \boldsymbol{\tau}, \boldsymbol{\nu}, \boldsymbol{\phi}_{\text{sr}}, \boldsymbol{\theta}_{\text{sr}}, \boldsymbol{\phi}_{\text{ris}} , \boldsymbol{\theta}_{\text{ris}} \right]^{\text{T}} \in \mathbb{R}^{8K + 2}, 
        \end{align}
        where 
        \begin{align*}
            \boldsymbol{\alpha}_{\text{Re}} &= \left[\text{Re}(\alpha_{1}), \cdots, \text{Re}(\alpha_{K})\right] \in \mathbb{C}^{K \times 1}, \\
            \boldsymbol{\alpha}_{\text{Im}} &= \left[\text{Im}(\alpha_{1}), \cdots, \text{Im}(\alpha_{K})\right] \in \mathbb{C}^{K \times 1}, \\
            \boldsymbol{\tau} &= [\tau_{1}, \cdots, \tau_{K}] \in \mathbb{C}^{K \times 1}, \\
            \boldsymbol{\nu} &= [\nu_{1}, \cdots, \nu_{K}] \in \mathbb{C}^{K \times 1}, \\
            \boldsymbol{\phi}_{\text{sr}} &= [\phi^{1}_{\text{sr}}, \cdots, \phi^{K}_{\text{sr}}] \in \mathbb{C}^{K \times 1}, \\
            \boldsymbol{\theta}_{\text{sr}} &= [\theta^{1}_{\text{sr}}, \cdots, \theta^{K}_{\text{sr}}] \in \mathbb{C}^{K \times 1}, \\
            \boldsymbol{\phi}_{\text{ris}} &= [\phi_{\text{ris}_{\text{A}}}, \phi^{1}_{\text{ris}_{\text{D}}}, \cdots, \phi^{K}_{\text{ris}_{\text{D}}}] \in \mathbb{C}^{K + 1 \times 1}, \\
            \boldsymbol{\theta}_{\text{ris}} &= [\theta_{\text{ris}_{\text{A}}}, \theta^{1}_{\text{ris}_{\text{D}}}, \cdots, \theta^{K}_{\text{ris}_{\text{D}}}] \in \mathbb{C}^{K + 1 \times 1}.
        \end{align*}
        Thus, the likelihood function is defined as \cite{Asim_2021}
        \begin{align}
            \begin{split}
                L(\boldsymbol{Y};\boldsymbol{\eta}) &= \frac{1}{\pi^{T M Q L_{\text{SR}}} \text{det}(\boldsymbol{R})} \\ &\text{exp}\left\{\hspace{-0.10cm}-\text{vec}\hspace{-0.05cm}\left(\boldsymbol{Y} \hspace{-0.10cm}-\hspace{-0.10cm} \boldsymbol{V}(\boldsymbol{\eta})\right)^{\text{H}} \hspace{-0.10cm}\boldsymbol{R}^{-1} \text{vec}\hspace{-0.05cm}\left(\boldsymbol{Y} \hspace{-0.10cm}-\hspace{-0.10cm} \boldsymbol{V}(\boldsymbol{\eta})\right)\hspace{-0.10cm}\right\}, \label{eq:likelihood_function} 
            \end{split}
        \end{align}
        with
            \begin{align} 
                \notag \boldsymbol{V}(\boldsymbol{\eta}) = \sum^{K}_{k = 1} &\{ \left[ \boldsymbol{b}_{\text{rx}}(\phi_{\text{ris}_{\text{A}}},\theta_{\text{ris}_{\text{A}}}) \boldsymbol{a}^{\text{T}}_{\text{st}}(\phi_{\text{st}},\hspace{-0.05cm}\theta_{\text{st}}) \boldsymbol{X} \text{D}(\boldsymbol{c}(\tau_{k}) \otimes \boldsymbol{d}(\nu_{k}))\right]^{\text{T}} \\ & \otimes \left[\alpha_{k} \boldsymbol{a}_{\text{sr}}(\phi^{k}_{\text{sr}},\theta^{k}_{\text{sr}}) \boldsymbol{b}^{\text{T}}_{\text{tx}}(\phi^{k}_{\text{ris}_{\text{D}}}, \theta^{k}_{\text{ris}_{\text{D}}})\right]\} \boldsymbol{S}. 
            \end{align}
        
        From (\ref{eq:likelihood_function}), we define the log-likelihood as $l(\boldsymbol{J};\boldsymbol{\eta}) = \text{ln}(L(\boldsymbol{J};\boldsymbol{\eta}))$, and we have $\sqrt{\text{CRLB}(\hat{\eta}_{i})} \geq \sqrt{[\boldsymbol{F}^{-1}(\boldsymbol{\eta})]_{ii}}$,
        where $[\boldsymbol{F}(\boldsymbol{\eta})]_{ij} = \mathbb{E}\left\{\frac{\partial l(\boldsymbol{Y};\boldsymbol{\eta})}{\partial \eta_{i}} \frac{\partial l(\boldsymbol{Y};\boldsymbol{\eta})}{\partial \eta_{j}}\right\}$. For the Gaussian case considered here, the FIM is given by \cite{Asim_2021}
        \begin{align}
            \notag [\boldsymbol{F}(\boldsymbol{\eta})]_{ij} &= 2 \text{Re}\left(\frac{\partial \text{vec}(\boldsymbol{V}(\boldsymbol{\eta}))^{\text{H}}}{\partial \eta_{i}} \boldsymbol{R}^{-1} \frac{\partial \text{vec}(\boldsymbol{V}(\boldsymbol{\eta}))}{\partial \eta_{j}}\right) \\ 
            &\hspace{+0.45cm}+ \text{tr}\left(\boldsymbol{R}^{-1} \frac{\partial \boldsymbol{R}}{\partial \eta_{i}} \boldsymbol{R}^{-1} \frac{\partial \boldsymbol{R}}{\partial \eta_{j}}\right),
        \end{align}
        where $\boldsymbol{R} = \mathbb{E}\left\{\text{vec}(\boldsymbol{Z}) \text{vec}(\boldsymbol{Z})^{\text{H}}\right\} = \sigma^{2}_{\boldsymbol{Z}} \boldsymbol{I}_{T M Q L_{\text{SR}}}$. If we apply property $\text{vec}^{\text{H}}(\boldsymbol{A}) \text{vec}(\boldsymbol{B}) = \text{tr}(\boldsymbol{A}^{\text{H}} \boldsymbol{B})$ \cite[p.~60]{petersen2008matrix}, then 
        \begin{align}
            [\boldsymbol{F}(\boldsymbol{\eta})]_{ij} &= \frac{2}{\sigma^{2}_{\boldsymbol{Z}}} 
            \text{Re}\left(\text{tr}\left(\frac{\partial \boldsymbol{V}(\boldsymbol{\eta})^{\text{H}}}{\partial \eta_{i}} \frac{\partial \boldsymbol{V}(\boldsymbol{\eta})}{\partial \eta_{j}}\right)\right). \label{eq:fim_matrix_block}
        \end{align}
        Detailed expressions for the partial derivatives in \eqref{eq:fim_matrix_block} are given in Appendix \ref{appendix_a}. 
        \begin{table}[!ht] 
            \centering
            \caption{Simulation parameters.}
            \label{tab:parameters}
            \resizebox{0.875\columnwidth}{!}{
            \begin{tabular}{|c|c|}
                \hline
                \acs{RIS} elements & $N_{y} N_{z} = 4 \times 4 = 16$ \\ \hline
                Antennas at the \ac{ST} & $L_{\text{ST}_{y}} L_{\text{ST}_{z}} = 2 \times 2 = 4$ \\ \hline
                Antennas at the \ac{SR} & $L_{\text{SR}_{y}} L_{\text{SR}_{z}} = 4 \times 4 = 16$ \\ \hline
                Arrays spacing & $\lambda/2$ \\ \hline
                Angles & $\mathcal{U}(0^{\circ}, 90^{\circ})$ \\ \hline
                \acs{ST} - \acs{RIS} & $\mathcal{U}(10\text{m},250\text{m})$ \\ \hline
                \acs{RIS} - Cluster & $\mathcal{U}(10\text{m},250\text{m})$\\ \hline
                Cluster - \acs{SR} & $\mathcal{U}(10\text{m},250\text{m})$ \\ \hline
                Velocity & $\mathcal{U}(-25\text{m/s},+25\text{m/s})$ \\ \hline
                \acs{RCS} & $2 \text{m}^{2}$ \\ \hline
                Symbol duration & $1/\Delta f$ \\ \hline
                Carrier frequency & $28$ GHz \\ \hline
                Subcarrier spacing $\Delta f$ & $120$ KHz \\ \hline
                Wavelength & $1.07 \times 10^{-2}$ m \\ \hline
                $i_{\text{max}}$ & $500$ \\ \hline
                $\delta$ & $10^{-6}$ \\ \hline
                Targets $K$ & $2$ \\ \hline
                Symbols $M$ & $8$ \\ \hline
                Subcarriers $Q$ & $8$ \\ \hline
                Time-slots $T$ & $512$ \\ \hline
            \end{tabular}}
        \end{table}
        
    \section{Numerical Results} \label{sec:numerical_results}
        In this section we evaluate the performance of the proposed \ac{TenDAE} framework for multi-target sensing using fully connected \ac{BD-RIS}. The results assess estimation accuracy, identifiability behavior, and performance--complexity trade-offs, and compare the proposed approach with diagonal-RIS-based counterparts and recent state-of-the-art methods \cite{kemal2024ris,ercan2025ris}.
        We consider a bistatic OFDM sensing scenario assisted by a fully-connected BD-RIS composed of $N = N_y \times N_z$ reflecting elements. The \ac{ST} and \ac{SR} employ uniform planar arrays with half-wavelength spacing. Targets are assumed to lie in the far field of both the \ac{BD-RIS} and the SR. The \ac{BD-RIS} phase-shift matrices are generated as random unitary matrices and remain constant over blocks of $MQ$ OFDM symbols. The pilot matrix $\boldsymbol{X}$ is constructed as a Hadamard matrix, and all target angles are drawn from a uniform distribution between $(0^{\circ},90^{\circ})$. The channel and parameter estimation accuracies are measured by the \ac{NMSE} and \ac{RMSE}:
        \begin{align}
            \text{NMSE}(\boldsymbol{X}) &= \mathbb{E}\left\{\frac{\left|\left|\boldsymbol{X} - \hat{\boldsymbol{X}}\right|\right|^{2}_{\text{F}}}{\left|\left|\boldsymbol{X}\right|\right|^{2}_{\text{F}}}\right\}, \\
            \text{RMSE}(\boldsymbol{x}) &= \sqrt{\mathbb{E}\left\{\left|\left|\boldsymbol{x} - \hat{\boldsymbol{x}}\right|\right|^{2}_{2}\right\}},
        \end{align}
        where the generic variable $\boldsymbol{X}$ represents a channel matrix and $\boldsymbol{x}$ a parameter vector. 
        We define $\text{SNR} = ||\mathcal{\boldsymbol{Y}}||^{2}_{\text{F}}\sigma^{2}_{\mathcal{\boldsymbol{Z}}}/||\mathcal{\boldsymbol{Z}}||^{2}_{\text{F}}$, where $\sigma^{2}_{\mathcal{\boldsymbol{Z}}}$ is the noise variance. The delay and Doppler parameters are normalized by the symbol period, $T_{s}$, to ensure fair performance comparisons. Unless otherwise stated, we use the system parameters summarized in Table \ref{tab:parameters}. All reported results are averaged over $M = 5000$ Monte Carlo realizations.
        
        To evaluate the performance of \ac{TenDAE}, we consider the following benchmarks: I) Maximum likelihood (ML)- and \ac{OMP}-based solutions based on \cite{ercan2025ris}, which offer distinct solutions for single-target and multi-target scenarios. For the ML solution, which is used for the single-target case or to initialize \ac{OMP}, the authors of \cite{ercan2025ris} solve a series of 4 sequential optimization problems where the last step depends on a 4D ML optimization. For \ac{OMP}, the detection problem is solved by sequentially estimating the target with the strongest signature, isolating each target's parameters one by one. II) \ac{HOSVD}-based Solutions: We adapt the single-target subspace-based approach from our previous work \cite{benicio2024low} to the multi-target case by leveraging the modular \ac{TenDAE} framework. III) \acf{CRLB}: We compute the \ac{CRLB} to establish the RMSE lower bound for the proposed bistatic \ac{MIMO} multi-target scenario.
        \begin{figure}
            \begin{minipage}{0.475\columnwidth}
                \includegraphics[width=0.9675\textwidth]{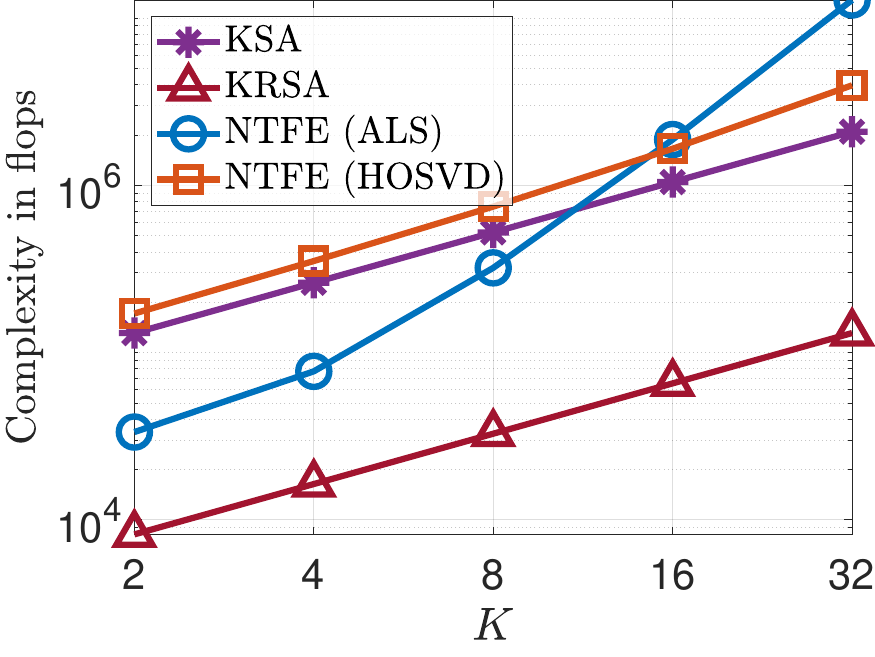}
                \caption*{(a) Number of targets $K$.}
            \end{minipage}
            \hfill
            \begin{minipage}{0.475\columnwidth} 
                \includegraphics[width=\textwidth]{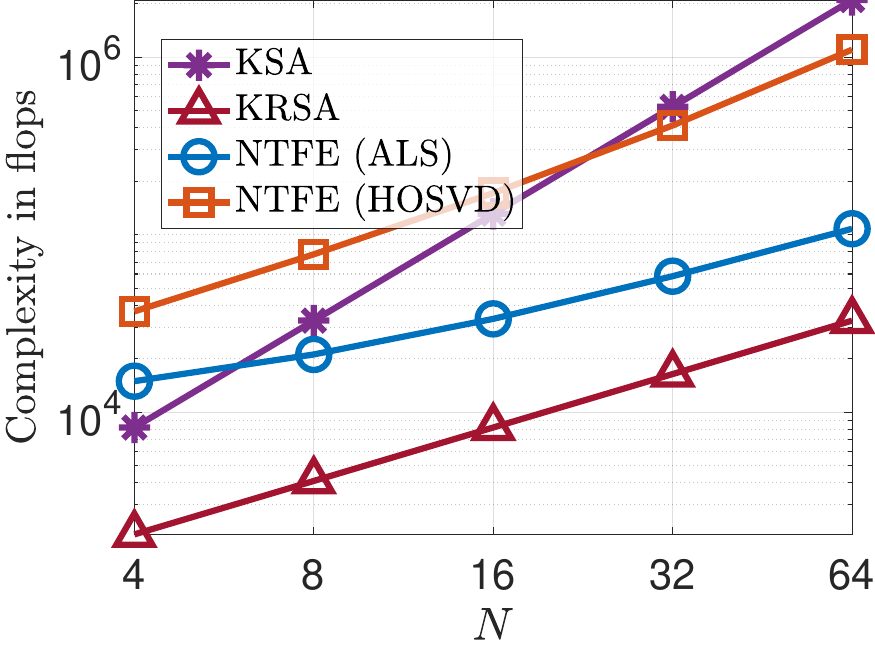}
                \caption*{(b) Number of elements $N$.}
            \end{minipage}
            \caption{Complexity as a function of the parameters in Table~\ref{tab:complexity}.}
            \label{fig:complexity_case}
        \end{figure}
         \begin{figure}
            \centering
            \includegraphics[width=0.875\columnwidth]{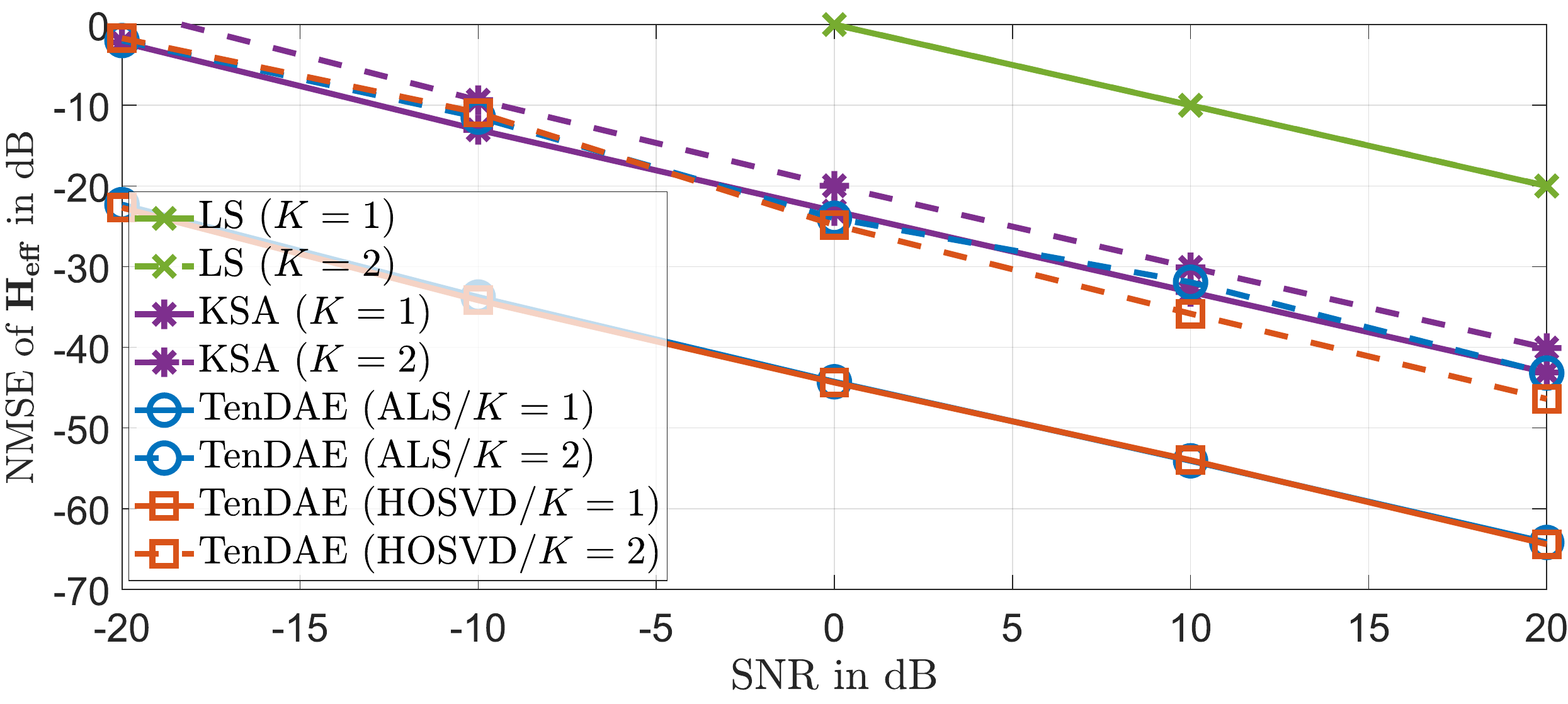}
            \caption{Channel estimation performance in terms of \ac{NMSE} as a function of the \ac{SNR}.}
            \label{fig:nmse_case_1}
        \end{figure} 

        \subsection{Complexity Evaluation}
In Fig.~\ref{fig:complexity_case}, we assess the computational complexity reported in Table~\ref{tab:complexity} as a function of key system parameters. This analysis highlights how the proposed algorithms scale with respect to the problem dimensions and clarifies the associated performance--complexity trade-offs intrinsic to the proposed scenario. To provide a conservative upper bound for the complexity of the iterative approach, we consider a worst-case scenario where convergence is achieved at a fixed threshold of $\text{ALS}_{\text{iter}} = 10$ iterations for both Fig.~\ref{fig:complexity_case}a and Fig.~\ref{fig:complexity_case}b. Typically, our iterative solution requires $2$ to $5$ iterations per tensor fit. Convergence requires more iterations at lower \ac{SNR} and fewer at higher \ac{SNR}, as \ac{ALS} convergence is strongly dependent on the noise level.
Fig.~\ref{fig:complexity_case}a shows the computational complexity as a function of the number of targets $K$. As expected, the overall complexity increases with $K$ for all methods. \ac{NTFE} \textit{via} \ac{ALS} exhibits the highest complexity for large $K$, since it jointly estimates the delay and Doppler parameters of all $K$ targets via two parallel tensor decompositions. This additional cost reflects the increased estimation capability enabled by the proposed framework, as it allows us to obtain information for all $K$ targets simultaneously without additional processing. Furthermore, Fig.~\ref{fig:complexity_case}b illustrates the complexity as a function of the number of \ac{BD-RIS} reflecting elements $N$. This result emphasizes the strong dependence of the first stage \ac{KSA} solution on $N$, despite its inability to determine the target parameters on its own. This behavior is consistent with the quadratic increase in complexity with $N$ for \ac{KSA} reported in Table~\ref{tab:complexity}. Both \ac{NTFE} algorithms scale linearly with $N$, although the \ac{HOSVD}-based solution tends to have a higher complexity due to the sequential \acp{SVD} that scales with $N$. Overall, these results confirm that the proposed algorithms offer a flexible trade-off between estimation quality and computational complexity, enabling system designers to tailor the sensing architecture to the available processing resources. For example, one could use our modular approach to the proposed scenario to combine the \ac{ALS} and \ac{HOSVD} solutions in a way that reduces the overall complexity of the estimation process, depending on the available sensing resources.
 
 \subsection{Channel Estimation Performance}
Figure~\ref{fig:nmse_case_1} evaluates the estimation accuracy of the effective channel, $\boldsymbol{H}_{\mathrm{eff}} = \sum_{k=1}^{K} \left\{ \boldsymbol{J}^{\mathrm{T}}_{k} \otimes \boldsymbol{G}_{k} \right\} \boldsymbol{S}$, as a function of the \ac{SNR}. The comparison includes the \ac{LS} estimator in~(\ref{eq:received_signal_filtered}), the \ac{KSA}-based approach in~(\ref{eq:ksa_1}), and the proposed parameter estimation framework. In this scenario, the \ac{LS} method estimates only the aggregated effective channel by right-filtering the received signal with the \ac{BD-RIS} phase-shift matrix in  (\ref{eq:received_signal_all_samples}), which is blind to the number of targets $K$, so both green curves share the same performance. The \ac{KSA} approach improves upon \ac{LS} by leveraging the Kronecker structure of the received signal, as shown in  (\ref{eq:received_signal_filtered}), enabling a more accurate channel reconstruction with a small degradation of around $1$~dB as the number of targets increases from 1 to 2. However, \ac{KSA} does not explicitly exploit the internal structure of the individual matrix factors that compose the sum of Kronecker products. In contrast, the proposed \ac{TenDAE} approach fully exploits the multidimensional tensor structure of the received sensing signal. By explicitly estimating the constituent factors $\{\boldsymbol{J}_{k}, \boldsymbol{G}_{k}\}_{k=1}^{K}$, \ac{TenDAE} achieves a more accurate reconstruction of the effective channel. As a result, it consistently outperforms both the \ac{LS} and \ac{KSA} estimators in terms of \ac{NMSE}, as observed in Fig.~\ref{fig:nmse_case_1}. These results demonstrate that exploiting the intrinsic tensor structure of the sensing model translates directly into tangible gains in channel estimation accuracy. In addition, it also provides direct access to the target parameter estimates.
            \begin{figure}
                \begin{minipage}{0.475\columnwidth}
                    \includegraphics[width=\textwidth]{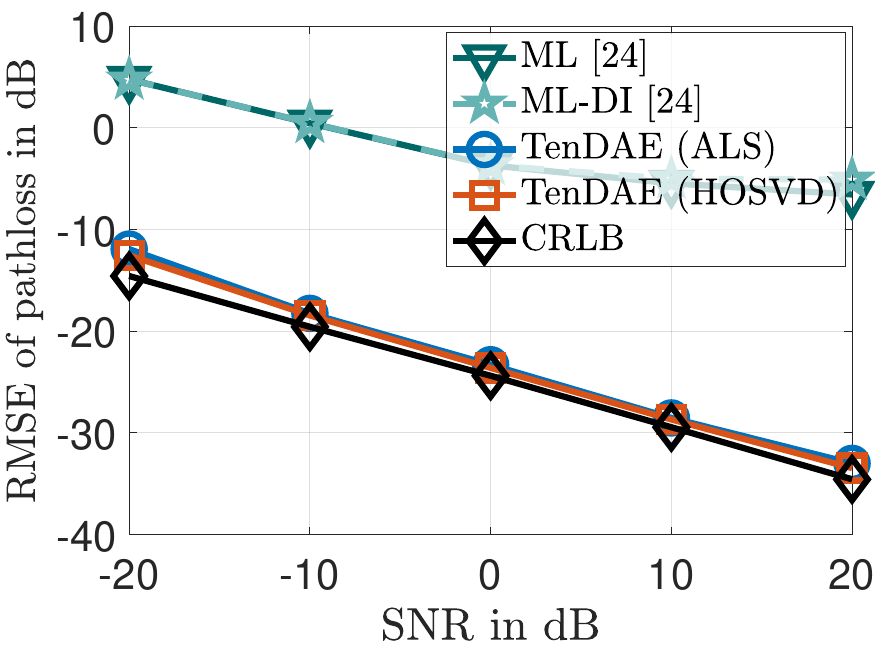}
                    \caption*{(a) \ac{RMSE} of the channel gain vs. \ac{SNR} at the SR.}  
                \end{minipage}
                \hfill
                \begin{minipage}{0.475\columnwidth}  
                    \includegraphics[width=\textwidth]{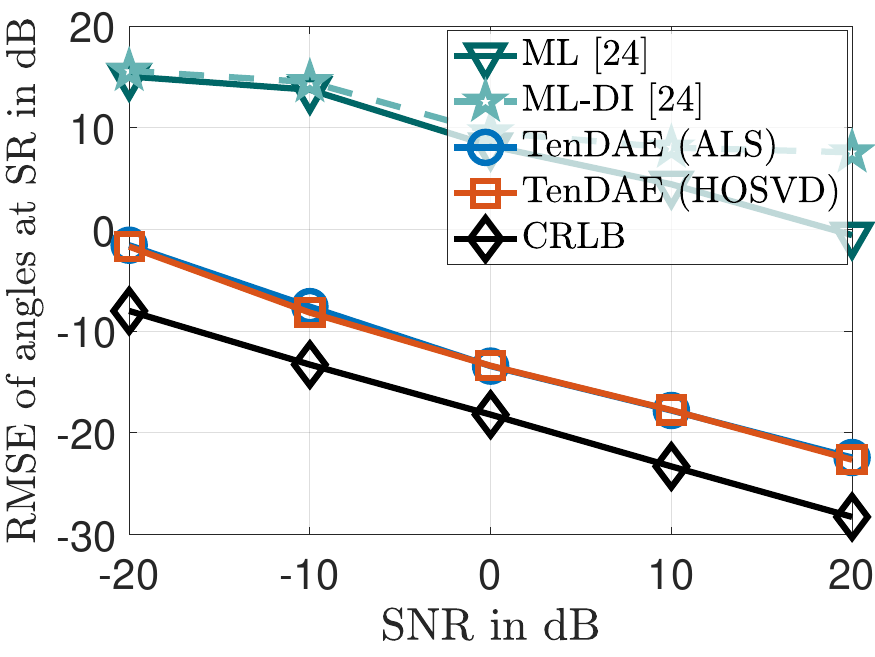}
                    \caption*{(b) \ac{RMSE} of the target angle vs. \ac{SNR}.}   
                \end{minipage}
                \vskip\baselineskip
                \begin{minipage}{0.475\columnwidth}   
                    \includegraphics[width=\textwidth]{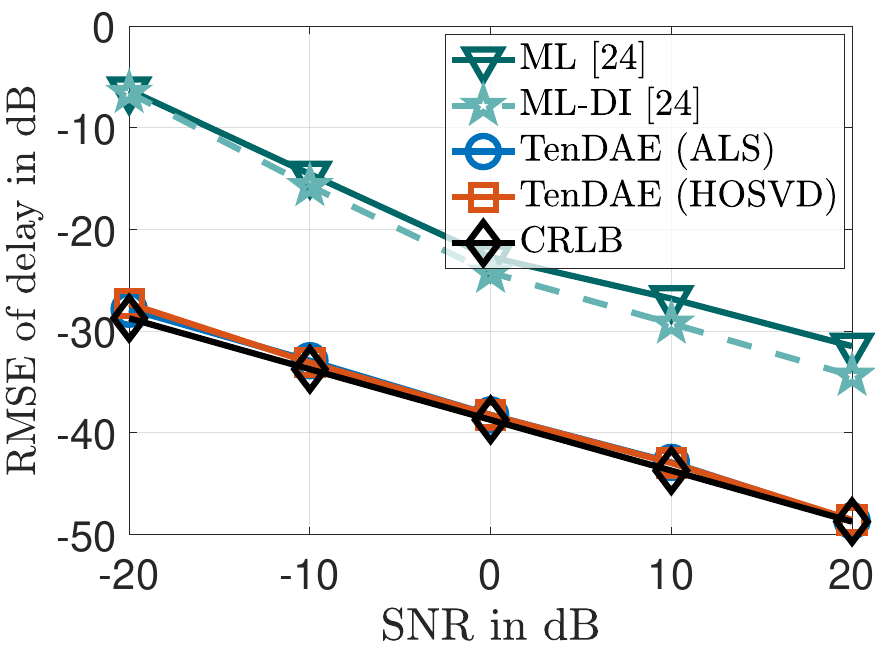}
                    \caption*{(c) Delay \ac{RMSE}  vs. \ac{SNR}.}
                \end{minipage}
                \hfill
                \begin{minipage}{0.475\columnwidth}   
                    \includegraphics[width=\textwidth]{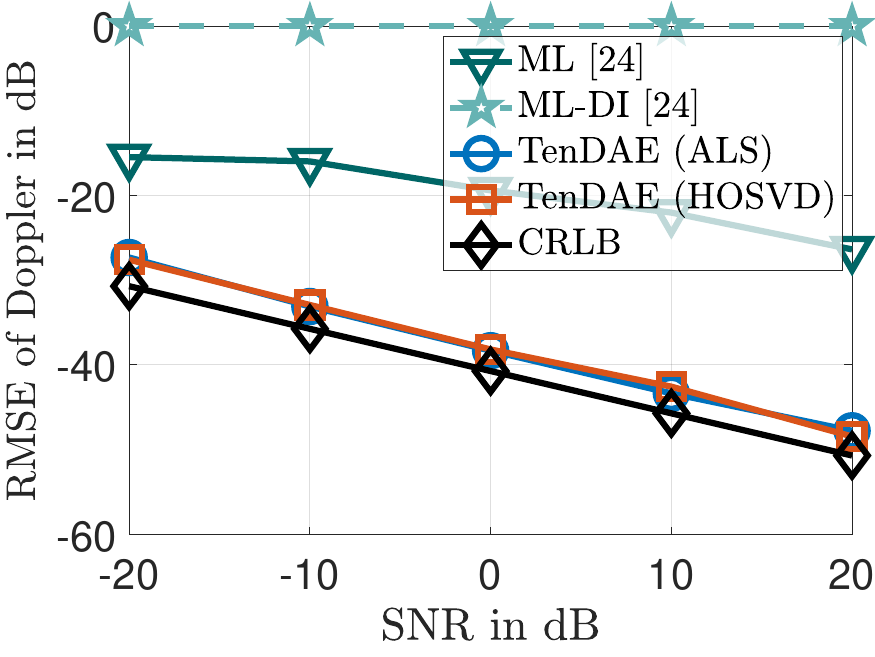}
                    \caption*{(d) Doppler \ac{RMSE} vs. \ac{SNR}.}  
                \end{minipage}
                \caption{\ac{RMSE} performance of the proposed solution in Alg. \ref{alg:proposed_1} and Alg. \ref{alg:proposed_2} in estimating the target parameters for both single and multi-target scenarios.} 
                \label{fig:rmse_case_5_single}
            \end{figure}

            \begin{figure}
                \begin{minipage}{0.475\columnwidth}
                    \includegraphics[width=\textwidth]{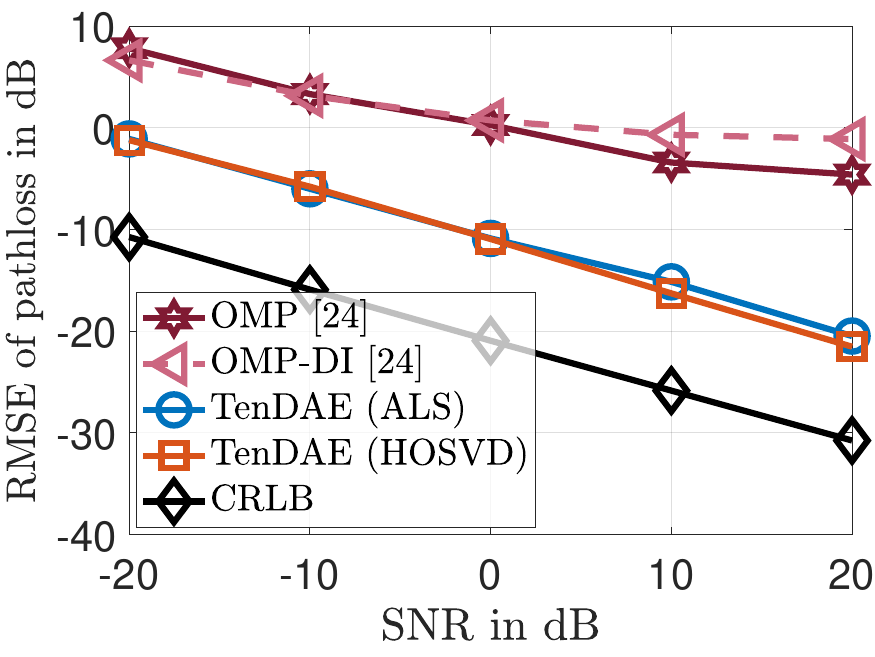}
                    \caption*{(a) \ac{RMSE} of the channel gain vs. \ac{SNR} at the \ac{SR}.}  
                \end{minipage}
                \hfill
                \begin{minipage}{0.475\columnwidth}  
                    \includegraphics[width=\textwidth]{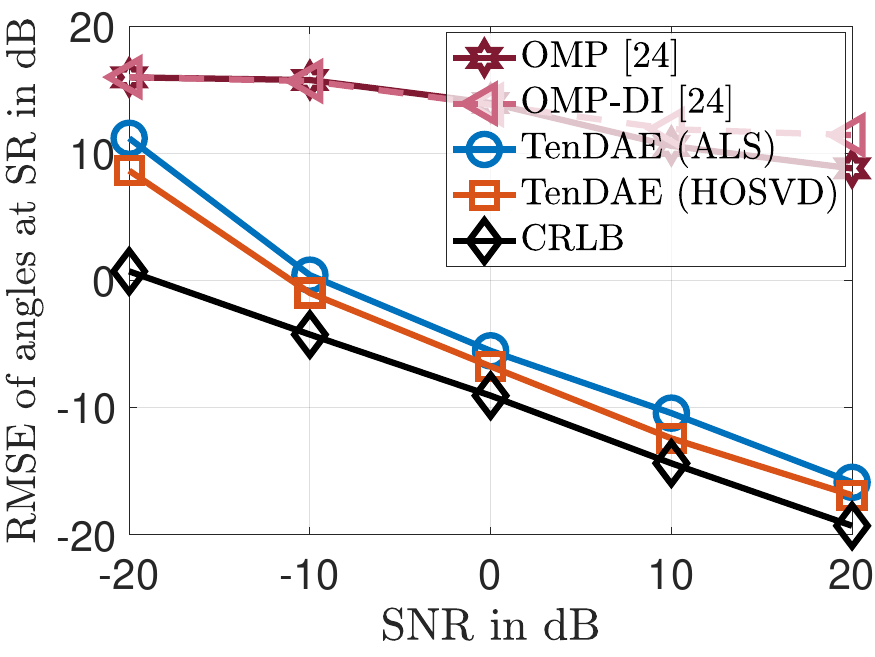}
                    \caption*{(b) \ac{RMSE} of the target angle vs.\ac{SNR}.}   
                \end{minipage}
                \vskip\baselineskip
                \begin{minipage}{0.475\columnwidth}   
                    \includegraphics[width=\textwidth]{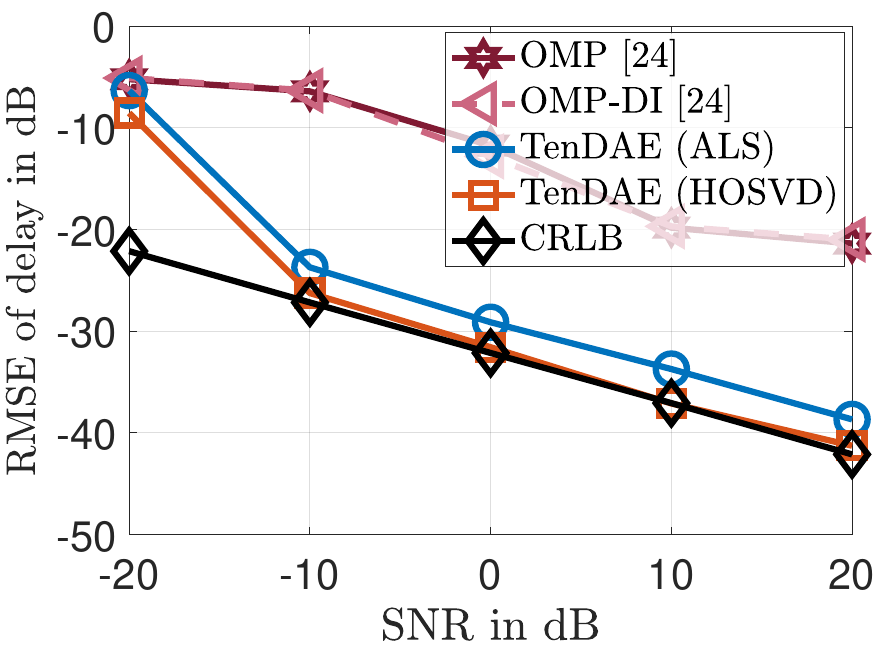}
                    \caption*{(c) Delay \ac{RMSE}  vs. \ac{SNR}.}
                \end{minipage}
                \hfill
                \begin{minipage}{0.475\columnwidth}   
                    \includegraphics[width=\textwidth]{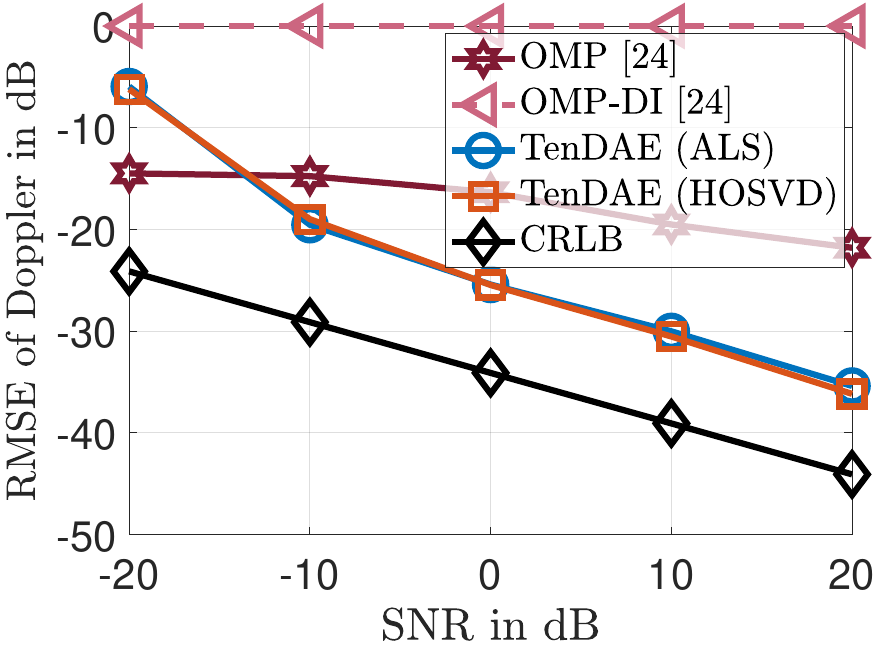}
                    \caption*{(d) Doppler \ac{RMSE} vs. \ac{SNR}.}  
                \end{minipage}
                \caption{\ac{RMSE} performance of the proposed solution in Alg. \ref{alg:proposed_1} and Alg. \ref{alg:proposed_2} in estimating the target parameters for both single- and multi-target scenarios.} 
                \label{fig:rmse_case_5_multi}
            \end{figure}
            
            \subsection{Parameter Estimation Performance}
                \begin{figure} 
                    \begin{minipage}{0.475\columnwidth}
                        \includegraphics[width=\textwidth]{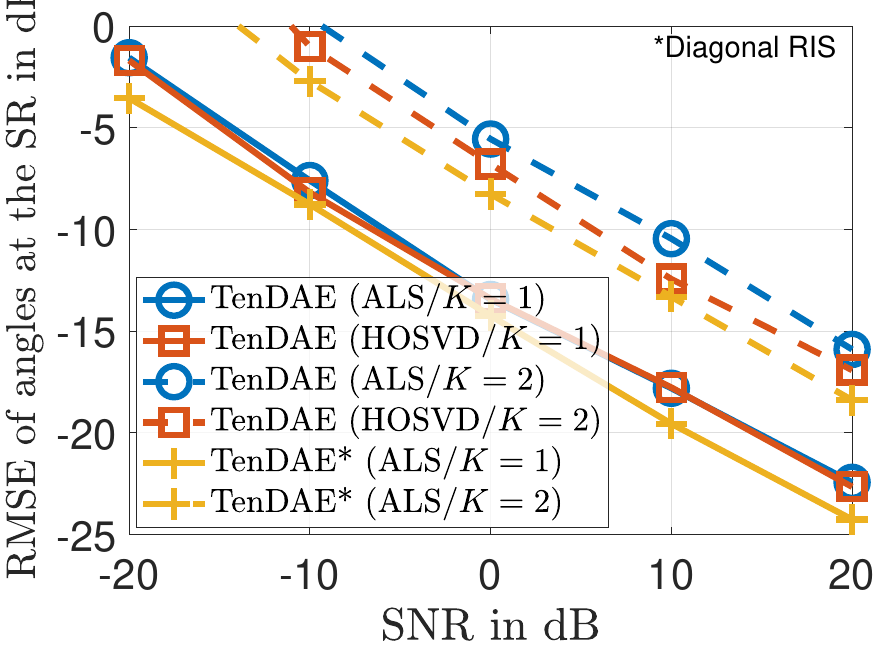}
                        \caption*{(a) \ac{RMSE} of the complex channel coefficients vs. \ac{SNR}.} 
                    \end{minipage}
                    \hfill
                    \begin{minipage}{0.475\columnwidth}  
                        \includegraphics[width=\textwidth]{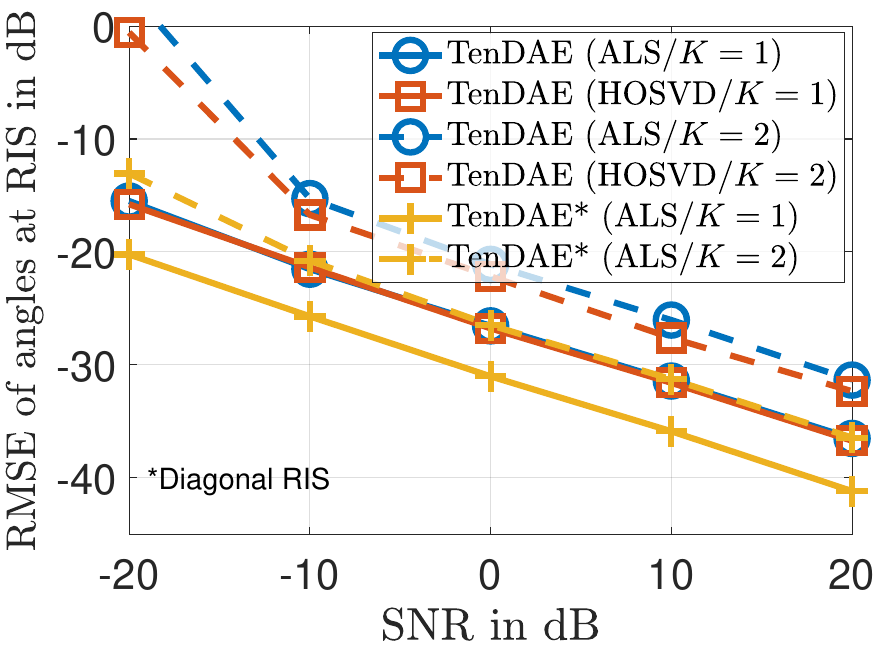}
                        \caption*{(b) \ac{RMSE} of the target angle vs. \ac{SNR}.}    
                    \end{minipage}
                    \vskip\baselineskip
                    \begin{minipage}{0.475\columnwidth}   
                        \includegraphics[width=\textwidth]{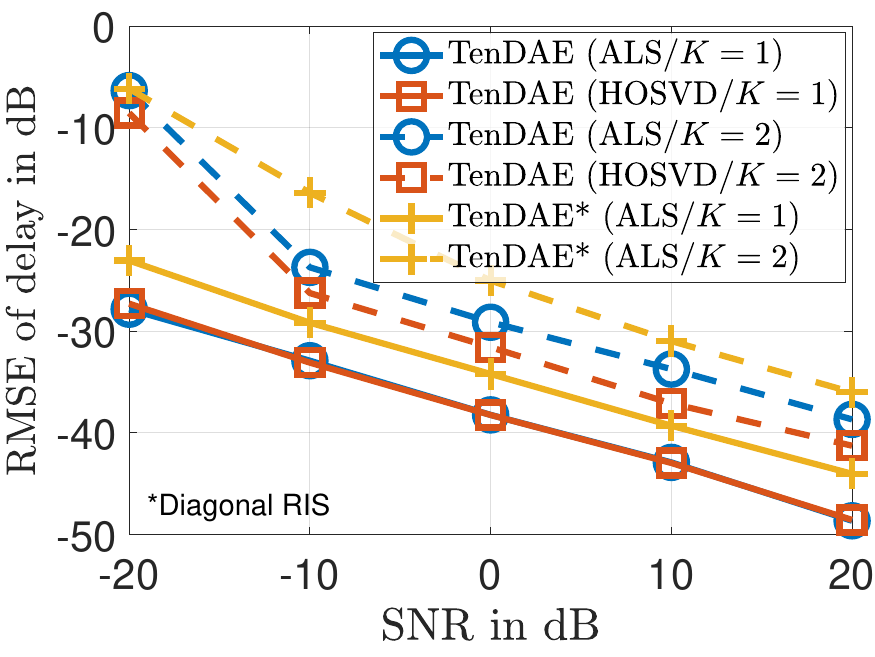}
                        \caption*{(c) Delay \ac{RMSE} vs. \ac{SNR}.} 
                    \end{minipage}
                    \hfill
                    \begin{minipage}{0.475\columnwidth}  
                        \includegraphics[width=\textwidth]{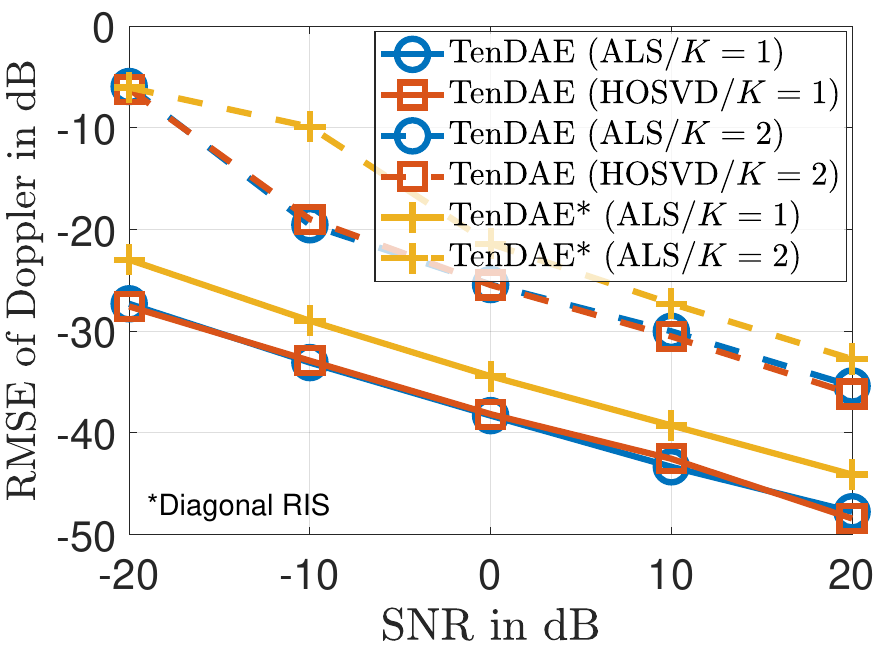}
                        \caption*{(d) Doppler \ac{RMSE} vs. \ac{SNR}.}   
                    \end{minipage}
                    \caption{Single-target RMSE comparison between the proposed solutions in Alg. \ref{alg:proposed_1} and Alg. \ref{alg:proposed_2} and the benchmark in \cite{ercan2025ris}.} 
                    \label{fig:rmse_case_6}
                \end{figure}
                
                In Fig.~\ref{fig:rmse_case_5_single}, we evaluate the parameter estimation performance in terms of \ac{RMSE} as a function of the \ac{SNR} for a single-target ($K = 1$) scenario, where the proposed \ac{TenDAE} framework—utilizing either \ac{ALS} or \ac{HOSVD}—closely follows the \ac{CRLB}. The gap with respect to the bound is less than $1$~dB for the channel coefficient, delay, and Doppler parameters, while a larger deviation of approximately $10$~dB is observed for angle estimation. This behavior is primarily attributed to the increased sensitivity of the angle parameters to noise and error propagation stemming from the use of \ac{KSA} in the first stage. To showcase the advantages of our model, we compare it against two benchmarks from the solution in \cite{ercan2025ris}. First, our \ac{TenDAE} framework outperforms the standard 4D ML approach by nearly $20$~dB across all parameters. This occurs because the solution in \cite{ercan2025ris} relies heavily on a precise initialization in the delay-Doppler domain, which requires significant sensing resources (e.g., $Q = 256$ and $M = 1120$). Conversely, \ac{TenDAE} achieves robust results with only $Q = 8$ subcarriers and $M = 8$ \ac{OFDM} symbols. Furthermore, the method in \cite{ercan2025ris} suffers from error propagation due to its use of four sequential ML optimizations and the inherent coupling between the Doppler and angular domains. Our framework, however, successfully decouples these domains by exploiting the nested-PARAFAC tensor $\boldsymbol{\mathcal{J}}$ in the second stage\footnote{See Fig. \ref{fig:solution_diagram}. After \ac{KSA} in the first stage, we decouple the angle and delay-Doppler parameters.}, which significantly improves performance with limited sensing resources. Finally, we include results for the \ac{DI} ML solution from \cite{ercan2025ris}. While this solution does not provide Doppler estimates, it otherwise follows the standard ML performance. This similarity arises because \cite{ercan2025ris} lacks sufficient resources for meaningful estimation in this specific scenario, although the \ac{DI}-based solution typically diverges further from the standard \ac{ML} approach as available sensing resources increase.

                Fig. \ref{fig:rmse_case_5_multi} illustrates that in the multi-target scenario ($K = 2$), the proposed solution continues to closely track the \ac{CRLB}, albeit with an increased gap ranging from $3$~dB to $10$~dB. The pathloss estimation experiences the largest degradation as it is the final step of the algorithm, where error propagation accumulates. Such performance loss is expected, as increasing the number of targets increases the dimensions of the parameter space. In the \ac{TenDAE} framework using \ac{ALS}, additional targets affect convergence behavior and increase computational complexity; more targets require additional iterations to satisfy the convergence criteria. The \ac{HOSVD}-based \ac{TenDAE} solution consistently outperforms the \ac{ALS} implementation in multi-target scenarios because it provides the optimal rank-$K$ approximation of the channel matrices for subsequent parameter extraction via \ac{ESPRIT}. The loss due to multiple targets can be mitigated by allocating additional sensing resources, including \ac{OFDM} symbols $M$, subcarriers $Q$, or time slots $T$. In comparison, the benchmark from \cite{ercan2025ris} employs \ac{OMP} with an intermediate single-target \ac{ML} step. Our solution outperforms this benchmark by $10$~dB in pathloss and Doppler estimation, and by nearly $20$~dB in the angle and delay domains. This improvement is expected, as the angle estimates in \cite{ercan2025ris} suffer from cascaded errors due to both the coarse delay-Doppler estimation and the initial \ac{ML} refinement.

                Fig.~\ref{fig:rmse_case_6} compares the proposed \ac{BD-RIS} architecture with a conventional diagonal \ac{RIS}. In Figs.~\ref{fig:rmse_case_6}a and~\ref{fig:rmse_case_6}b, we evaluate the \ac{RMSE} of both architectures for the target and RIS angle parameters. Unlike the BD-RIS, the system with a diagonal \ac{RIS} is unable to uniquely estimate arrival and departure angles asociated with the RIS. Thus, even without information beyond the angles estimated at the \ac{ST}, the BD-RIS-aided system can uniquely estimate the angles passing through the surface. The results in Fig.~\ref{fig:rmse_case_6}c and Fig.~\ref{fig:rmse_case_6}d indicate that the \ac{BD-RIS} consistently outperforms the diagonal \ac{RIS} in terms of \ac{RMSE} for delay and Doppler estimation. Comparisons of the \ac{ALS} and \ac{HOSVD} implementations of \ac{TenDAE} show similar behavior as those in Figs.~\ref{fig:rmse_case_5_single} and~\ref{fig:rmse_case_5_multi}. While the diagonal \ac{RIS} provides a marginal $1$~dB improvement in angle estimation at the \ac{SR} due to identical receiver array configurations, it suffers a performance loss of approximately $5$~dB in delay and Doppler estimation compared to the \ac{BD-RIS}.
                               
\begin{figure}
                    \begin{minipage}{0.475\columnwidth}
                        \includegraphics[width=\textwidth]{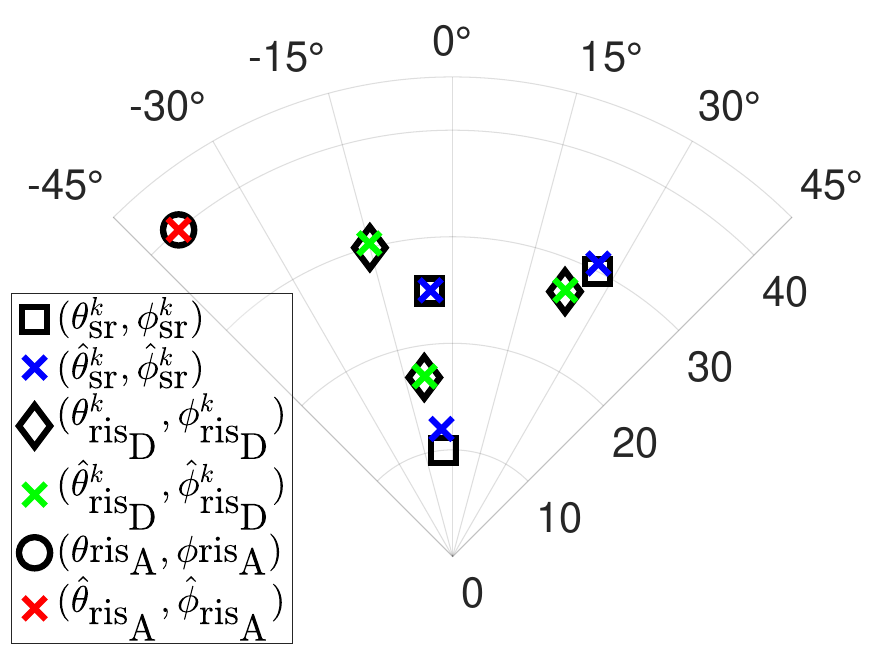}
                        \caption*{(a) Evaluation of the angle parameter estimates.} 
                    \end{minipage}
                    \hfill
                    \begin{minipage}{0.475\columnwidth}  
                        \includegraphics[width=\textwidth]{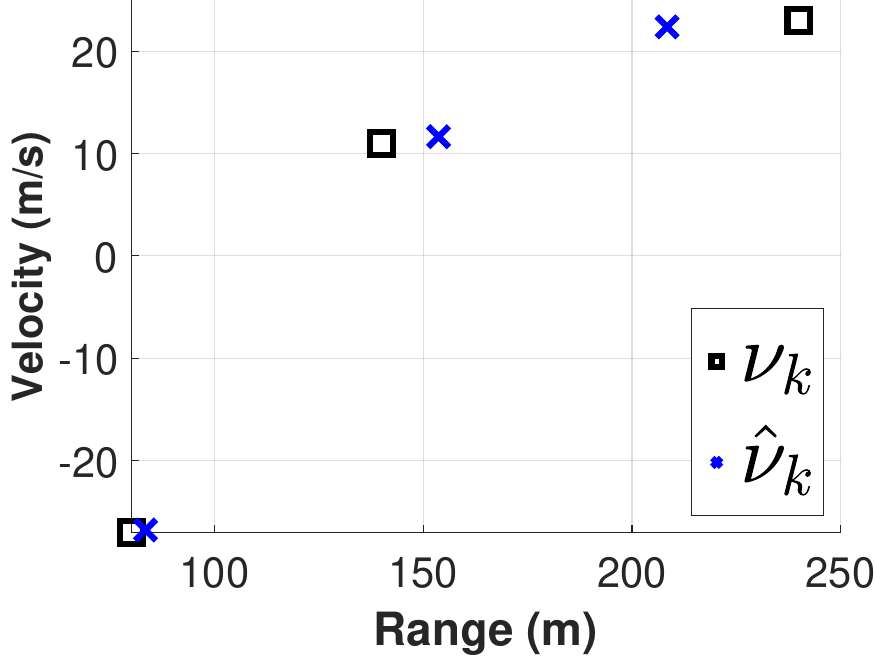}
                        \caption*{(b) Evaluation of velocity and range parameter estimates.} 
                    \end{minipage}
                    \caption{Evaluation of the proposed solution in Alg. \ref{alg:proposed_1} and Alg. \ref{alg:proposed_2} for a training \ac{SNR} of $15$dB and $K = 3$ targets.} 
                    \label{fig:scatter_case}
                \end{figure}

                In Fig.~\ref{fig:scatter_case}, we present a visual representation of scenario with $K=3$ targets and a training SNR of $15$~dB. In these plots, the \ac{AoA} at the \ac{SR} is denoted by black squares, the \ac{AoD} at the \ac{BD-RIS} by black diamonds, and the \ac{AoA} at the \ac{BD-RIS} by black circles. \ac{TenDAE} demonstrates robustness in estimating angles from diverse sources; for instance, even when the \ac{AoD} at the \ac{BD-RIS} is physically close to the \ac{AoA} at the \ac{SR}, both are accurately estimated. However, as expected, performance degradation occurs when the \ac{AoA} values at the \ac{SR} are close to one another. In the spatial domain, higher target velocities make it more difficult to obtain precise Doppler parameter estimates. Conversely, our framework exhibits lower sensitivity to range variations, yielding more reliable estimates for the delay parameter.

    \section{Conclusion}
        \label{sec:conclusion}
        We have proposed \ac{TenDAE}, a two-stage framework for multi-target parameter estimation that leverages the intrinsic multidimensional structure of the sensing signal received from a BD-RIS to solve tensor approximation problems involving either closed-form \ac{HOSVD} or iterative \ac{ALS}-based solutions. The proposed approach decomposes the sensing task into a sequence of modular stages, namely \ac{KSA} and \ac{NTFE}, followed by high-resolution parameter extraction, enabling joint estimation of delay, Doppler, and angle parameters with low-complexity processing. By exploiting the Kronecker structure induced by fully connected \ac{BD-RIS} architectures, the \ac{TenDAE} enables unique estimation of the arrival and departure angles at the BD-RIS. Our results demonstrated that \ac{TenDAE} consistently outperforms state-of-the-art benchmarks in \ac{RMSE} while exhibiting favorable, controllable trade-offs among estimation accuracy, computational complexity, and sensing resources. In particular, the modular \ac{KSA}-\ac{NTFE} decomposition provides a flexible design paradigm, allowing system designers to balance performance and complexity according to the available sensing resources. Also, numerical results show that our proposed solution outperforms current state-of-the-art methods \cite{kemal2024ris,ercan2025ris} in terms of \ac{RMSE} of the estimated parameters. Future work will investigate alternative realizations of the \ac{TenDAE} framework that avoid explicit right-filtering of the \ac{BD-RIS} phase-shift matrix, thereby mitigating the computational burden associated with the initial \ac{LS} step.

    \section*{Acknowledgments}
        We gratefully acknowledge the authors of \cite{ercan2025ris} for kindly providing the necessary code, which enabled a transparent and equitable comparison with the results presented herein.
    \appendices
    \section{Derivation of the FIM} \label{appendix_a}
         We first define the necessary expressions for the partial derivatives with respect to $\text{Re}(\boldsymbol{\alpha})$ and $\text{Im}(\boldsymbol{\alpha})$, where $\boldsymbol{\alpha} = \left[\alpha_{1}, \cdots, \alpha_{K}\right]$, which are calculated as
        \begin{align}
            \frac{\partial \alpha_{k}}{\partial \text{Re}(\alpha_{i})} &= \frac{\partial \alpha_{k}^{*}}{\partial \text{Re}(\alpha_{i})} = 
            \begin{cases}
                1 \text{ if } (k = i) \\
                0 \text{ if } (k \neq i),
            \end{cases} \\
            \frac{\partial \alpha_{k}}{\partial \text{Im}(\alpha_{i})} &= \frac{\partial \alpha_{k}^{*}}{\partial \text{Im}(\alpha_{i})} = 
            \begin{cases}
                j \text{ if } (k = i) \\
                0 \text{ if } (k \neq i).
            \end{cases}
        \end{align}
        The response of the $n$th element of the steering vector of a uniform rectangular array is given by
        \begin{align}
            \left[\boldsymbol{a}(\phi, \theta)\right]_{n} = e^{-j\left[(n_{y} - 1) \pi \sin(\phi) \sin(\theta) + (n_{z} - 1) \pi \cos(\phi)\right]}
        \end{align}
        where $n = n_{z} + (n_{y} - 1)N_{y}$, $n_{y} \in \{1, \cdots, N_{y}\}$, $n_{z} \in \{1, \cdots, N_{z}\}$, and $N = N_{y} N_{z}$. The partial derivatives of $\boldsymbol{a}(\phi, \theta)$ are calculated as in \cite{Asim_2021}:
        \begin{align}
            \notag &\frac{\partial \boldsymbol{a}(\phi, \theta)}{\partial \phi} \hspace{-0.1cm}=\hspace{-0.1cm} \left\{-j \pi \left[(n_{y} - 1) \cos(\phi) \sin(\theta) - (n_{z} - 1)\sin(\phi)\right] \right. \\ &\hspace{+0.475cm}\left.\left[\boldsymbol{a}(\phi, \theta)\right]_{n}\right\}^{N_{y} N_{z}}_{n = 1}, \\
            &\frac{\partial \boldsymbol{a}(\phi, \theta)}{\partial \theta} \hspace{-0.1cm}=\hspace{-0.1cm} \{-j \pi (n_{y} \hspace{-0.1cm}-\hspace{-0.1cm} 1) \sin(\phi) \cos(\theta) \hspace{-0.05cm}\left[\boldsymbol{a}(\phi, \theta)\right]_{n}\hspace{-0.05cm}\}^{N_{y} N_{z}}_{n = 1}.
        \end{align}
        
        As an example, the entries $\left[\boldsymbol{F}_{\boldsymbol{\alpha}_{\text{Re}} \boldsymbol{\phi}_{\text{sr}}}\right]_{ij}$ of the FIM block corresponding to $\boldsymbol{\alpha}_{\text{Re}}$ and $\boldsymbol{\phi}_{\text{sr}}$ are expressed as 
        \begin{align}
            \left[\boldsymbol{F}_{\boldsymbol{\alpha}_{\text{Re}} \boldsymbol{\phi}_{\text{sr}}}\right]_{ij} = \frac{2}{\sigma^{2}_{\boldsymbol{Z}}} 
            \text{Re}\left(\text{tr}\left(\frac{\partial \boldsymbol{V}(\boldsymbol{\eta})^{\text{H}}}{\partial \text{Re}(\alpha_{i})} \frac{\partial \boldsymbol{V}(\boldsymbol{\eta})}{\partial \phi^{j}_{\text{sr}}}\right)\right),
        \end{align}
        where 
        \begin{align}
            \notag\frac{\partial \boldsymbol{V}(\boldsymbol{\eta})}{\partial \text{Re}(\alpha_{i})} &= \sum^{K}_{k = 1} \biggl\{ \boldsymbol{J}_{k}^{\text{T}} \otimes \frac{\partial \boldsymbol{\alpha}_{k}}{\partial \text{Re}(\alpha_{i})} \boldsymbol{a}_{\text{sr}}(\phi^{k}_{\text{sr}},\theta^{k}_{\text{sr}})\boldsymbol{b}^{\text{T}}_{\text{tx}}(\phi^{k}_{\text{ris}_{\text{D}}}, \theta^{k}_{\text{ris}_{\text{D}}})\biggr\} \boldsymbol{S}, \\
            \notag \frac{\partial \boldsymbol{V}(\boldsymbol{\eta})}{\partial \phi^{j}_{\text{sr}}} &= \sum^{K}_{k = 1} \bigg\{\boldsymbol{J}^{\text{T}}_{k} \otimes \boldsymbol{\alpha}_{k} \frac{\partial \boldsymbol{a}_{\text{sr}}(\phi^{k}_{\text{sr}},\theta^{k}_{\text{sr}})}{\partial \phi^{j}_{\text{sr}}} \boldsymbol{b}^{\text{T}}_{\text{tx}}(\phi^{k}_{\text{ris}_{\text{D}}}, \theta^{k}_{\text{ris}_{\text{D}}})\bigg\} \boldsymbol{S},
        \end{align}
        from which we compute  
         \begin{align}
            \frac{\partial \boldsymbol{\alpha}_{k}}{\partial \text{Re}(\alpha_{i})} &= 
            \begin{cases}
                1 \text{ if } (k = i) \\
                0 \text{ if } (k \neq i),
            \end{cases} \\
            \notag \frac{\partial \boldsymbol{a}_{\text{sr}}(\phi^{k}_{\text{sr}},\theta^{k}_{\text{sr}})}{\partial \phi^{j}_{\text{sr}}} &= \{-j \pi [(l_{\text{SR}y} - 1) \cos(\phi^{k}_{\text{sr}}) \sin(\theta^{k}_{\text{sr}}) - \\ &(l_{\text{SR}z} - 1)\sin(\phi^{k}_{\text{sr}})] \left[\boldsymbol{a}(\phi^{k}_{\text{sr}}, \theta^{k}_{\text{sr}})\right]_{n}\}^{L_{\text{SR}y} L_{\text{SR}z}}_{l_{\text{SR}} = 1}.
        \end{align}

        The remaining blocks of the \ac{FIM} are acquired similarly using  (\ref{eq:fim_matrix_block}) as a guide to obtain the partial derivatives.
        
    \bibliographystyle{IEEEtran}
    \bibliography{IEEEexample}
    
\end{document}